\documentclass[fleqn,usenatbib]{mnras}

\usepackage{newtxtext,newtxmath}

\usepackage[T1]{fontenc}

\DeclareRobustCommand{\VAN}[3]{#2}
\let\VANthebibliography\thebibliography
\def\thebibliography{\DeclareRobustCommand{\VAN}[3]{##3}\VANthebibliography}

\usepackage{graphicx}	
\usepackage{amsmath}	
\usepackage{orcidlink}

\newcommand{\Mhalo}{{M}_\rmn{200c}}
\newcommand{\Rhalo}{{R}_\rmn{200c}}
\newcommand{\bs}[1]{\boldsymbol{#1}}

\title[Magnetic fields from dwarfs to groups]{Magnetic field amplification in cosmological zoom simulations from dwarf galaxies to galaxy groups}

\author[R. Pakmor et al.]{%
R\"udiger Pakmor$^1$\thanks{rpakmor@mpa-garching.mpg.de}\orcidlink{0000-0003-3308-2420},
Rebekka Bieri$^2$,
\orcidlink{0000-0002-4554-4488}
Freeke van~de~Voort$^3$
\orcidlink{0000-0002-6301-638X},
Maria Werhahn$^1$
\orcidlink{0000-0003-4984-4389}, 
Azadeh Fattahi$^{4}$
\orcidlink{0000-0002-6831-5215},
\newauthor Thomas Guillet$^{5}$
\orcidlink{0000-0002-0271-5953}, 
Christoph Pfrommer$^6$
\orcidlink{0000-0002-7275-3998}, 
Volker Springel$^1$
\orcidlink{0000-0001-5976-4599}, 
Rosie Y. Talbot$^1$
\orcidlink{0000-0001-9393-7879}
\vspace*{0.1cm}\\
$^{1}$Max-Planck-Institut f\"{u}r Astrophysik, Karl-Schwarzschild-Str. 1, D-85748, Garching, Germany\\
$^{2}$Institute for Computational Science, University of Zurich, Zurich, Switzerland\\
$^{3}$Cardiff Hub for Astrophysics Research and Technology, School of Physics and Astronomy, Cardiff University, Queen’s Buildings, Cardiff CF24 3AA, UK\\
$^{4}$Institute for Computational Cosmology, Department of Physics, Durham University, South Road, Durham DH1 3LE, UK \\
$^{5}$Physics and Astronomy, University of Exeter, Exeter EX4 4QL, UK\\
$^{6}$Leibniz-Institute f\"{u}r Astrophysik Potsdam (AIP), An der Sternwarte 16, 14482 Potsdam, Germany
}

\date{Accepted 2024 January 09. Received 2024 January 09; in original form 2023 September 22}

\pubyear{2024}

\begin{document}
\label{firstpage}
\pagerange{\pageref{firstpage}--\pageref{lastpage}}
\maketitle

\begin{abstract}
Magnetic fields are ubiquitous in the Universe. Recently, cosmological simulations of galaxies have successfully begun to incorporate magnetic fields and their evolution in galaxies and their haloes.
However, so far they have mostly focused on Milky Way-like galaxies. Here we analyse a sample of high resolution cosmological zoom simulations of disc galaxies in haloes with mass $\Mhalo$ from $10^{10}\,\rmn{M}_\odot$ to $10^{13}\,\rmn{M}_\odot$, simulated with the Auriga galaxy formation model.
We show that with sufficient numerical resolution the magnetic field amplification and saturation is converged. The magnetic field strength reaches equipartition with turbulent energy density for galaxies in haloes with $\Mhalo\gtrsim 10^{11.5}\,\mathrm{M_\odot}$. For galaxies in less massive haloes, the magnetic field strength saturates at a fraction of equipartition that decreases with decreasing halo mass. For our lowest mass haloes, the magnetic field saturates significantly below $10\%$ of equipartition. We quantify the resolution we need to obtain converged magnetic field strengths and discuss our resolution requirements also in the context of the IllustrisTNG cosmological box simulations.
We show that, at $z=0$, rotation-dominated galaxies in our sample exhibit for the most part an ordered large scale magnetic field, with fewer field reversals in more massive galaxies.
Finally, we compare the magnetic fields in our cosmological galaxies at $z=0$ with simulations of isolated galaxies in a collapsing halo setup.
Our results pave the way for detailed studies of cosmic rays and other physical processes in similar cosmological galaxy simulations that crucially depend on the strength and structure of magnetic fields.
\end{abstract}




\begin{table*}
\begin{center}
\begin{tabular}{ c c c c c c c c c c c c c }
\hline
    Name & $\Mhalo$ & $\Rhalo$ &    $M_*$ & $\dot{M}_{100\mathrm{Myr}}$ & $M_\mathrm{cell}$ & $r_\mathrm{disc}$ & $h_\mathrm{disc}$ & $B_\mathrm{disc}$ & $B_\mathrm{halo}$ & $\frac{E_\mathrm{disc,mag}}{E_\mathrm{disc,turb}}$ & $\frac{E_\mathrm{disc,mag,\phi}}{E_\mathrm{disc,mag,tot}}$\\
         & $[\rmn{M}_\odot]$ & $\mathrm{[kpc]}$ & $[\rmn{M}_\odot]$ & $[\rmn{M}_\odot / \mathrm{yr}]$ & $[\rmn{M}_\odot]$ & $\mathrm{[kpc]}$ & $\mathrm{[kpc]}$ & $[\mu G]$ & $[\mu G]$ &          &         \\
\hline
1e10\_h12 & $10^{9.9}$ & $40$ & $10^{7.7}$ & $1.1 \times 10^{-2}$ & $1 \times 10^{2}$ & $  0.9$ & $  0.6$ & $9.0 \times 10^{-1}$ & $4.3 \times 10^{-4}$ & $0.01$ & $0.3$ \\
1e10\_h8 & $10^{10.1}$ & $49$ & $10^{8.8}$ & $7.6 \times 10^{-2}$ & $1 \times 10^{2}$ & $  3.2$ & $  2.8$ & $1.2 \times 10^{0}$ & $1.2 \times 10^{-3}$ & $0.03$ & $0.3$ \\
1e10\_h11 & $10^{10.4}$ & $60$ & $10^{8.7}$ & $1.1 \times 10^{-1}$ & $1 \times 10^{2}$ & $  5.7$ & $  3.4$ & $1.2 \times 10^{0}$ & $3.1 \times 10^{-3}$ & $0.13$ & $0.3$ \\
1e10\_h9 & $10^{10.6}$ & $69$ & $10^{8.9}$ & $1.2 \times 10^{-1}$ & $8 \times 10^{2}$ & $  9.1$ & $  3.8$ & $10.0 \times 10^{-1}$ & $8.7 \times 10^{-3}$ & $0.13$ & $0.4$ \\
1e11\_h10 & $10^{10.9}$ & $91$ & $10^{9.6}$ & $5.6 \times 10^{-1}$ & $6 \times 10^{3}$ & $ 12.6$ & $  5.9$ & $1.8 \times 10^{0}$ & $2.9 \times 10^{-2}$ & $0.16$ & $0.4$ \\
1e11\_h11 & $10^{11.0}$ & $94$ & $10^{9.7}$ & $8.1 \times 10^{-1}$ & $6 \times 10^{3}$ & $ 10.1$ & $  5.4$ & $2.7 \times 10^{0}$ & $2.1 \times 10^{-2}$ & $0.24$ & $0.4$ \\
1e11\_h5 & $10^{11.4}$ & $138$ & $10^{10.1}$ & $3.8 \times 10^{-1}$ & $6 \times 10^{3}$ & $ 31.9$ & $  3.4$ & $1.6 \times 10^{0}$ & $6.6 \times 10^{-2}$ & $0.87$ & $0.6$ \\
1e11\_h4 & $10^{11.4}$ & $138$ & $10^{10.1}$ & $1.3 \times 10^{0}$ & $6 \times 10^{3}$ & $ 21.9$ & $  5.8$ & $2.8 \times 10^{0}$ & $3.6 \times 10^{-2}$ & $0.87$ & $0.5$ \\
1e12\_h12 & $10^{12.0}$ & $216$ & $10^{10.8}$ & $6.5 \times 10^{0}$ & $5 \times 10^{4}$ & $ 21.4$ & $  6.1$ & $6.1 \times 10^{0}$ & $1.3 \times 10^{-1}$ & $1.02$ & $0.7$ \\
1e12\_h5 & $10^{12.1}$ & $222$ & $10^{10.9}$ & $4.3 \times 10^{0}$ & $5 \times 10^{4}$ & $ 23.3$ & $  4.1$ & $5.3 \times 10^{0}$ & $1.2 \times 10^{-1}$ & $1.06$ & $0.7$ \\
1e13\_h4 & $10^{12.5}$ & $305$ & $10^{11.2}$ & $3.8 \times 10^{0}$ & $5 \times 10^{4}$ & $ 40.2$ & $  4.3$ & $4.9 \times 10^{0}$ & $2.1 \times 10^{-1}$ & $2.88$ & $0.9$ \\
1e13\_h3 & $10^{12.5}$ & $310$ & $10^{11.2}$ & $7.0 \times 10^{0}$ & $5 \times 10^{4}$ & $ 34.9$ & $  6.0$ & $4.6 \times 10^{0}$ & $1.6 \times 10^{-1}$ & $0.65$ & $0.7$ \\
1e13\_h8 & $10^{13.0}$ & $458$ & $10^{11.4}$ & $3.0 \times 10^{0}$ & $5 \times 10^{4}$ & $ 46.9$ & $  7.1$ & $2.8 \times 10^{0}$ & $1.6 \times 10^{-1}$ & $0.48$ & $0.9$ \\
1e13\_h7 & $10^{13.1}$ & $478$ & $10^{11.4}$ & $5.0 \times 10^{-2}$ & $5 \times 10^{4}$ & $  0.9$ & $  0.4$ & $2.1 \times 10^{1}$ & $1.2 \times 10^{-1}$ & $0.15$ & $0.7$ \\
\hline
\end{tabular}
\end{center}

\caption{Properties of the simulated galaxies at their highest resolution level at $z=0$. Columns show, from left to right, the name of the halo, its total mass $\Mhalo$ and size $\Rhalo$, its stellar mass $M_*$, the time averaged star formation rate $\dot{M}_{100\mathrm{Myr}}$ within the last $100\,\mathrm{Myr}$, the highest baryonic mass resolution $M_\mathrm{cell}$ we ran this halo at, the radius $r_\mathrm{disc}$ and height $h_\mathrm{disc}$ of the gas disc, that together define the volume of the disc, the average magnetic field strength in the disc $B_\mathrm{disc}$ and the full halo $B_\mathrm{halo}$, as well as the ratio between total magnetic and total turbulent energy in the volume of the disc and the fraction of magnetic energy in the azimuthal component of the magnetic field in the disc.}
\label{tab:galprops}
\end{table*}

\section{Introduction}

Our present-day Universe is almost completely ionised and permeated by magnetic fields \citep{Beck2015}. In the Milky Way and nearby disc galaxies magnetic fields are observed with various tracers to be consistent with equipartition between magnetic energy density and turbulent energy density \citep{Boulares1990, Beck1996, Beck2015}, though with significant systematic uncertainties.

For Milky Way-like galaxies we typically observe magnetic fields with strengths of order $1-10\,\mathrm{\mu G}$ \citep{Beck2013}. To reach such magnetic field strengths an efficient amplification of the magnetic fields over many orders of magnitude starting from a physically plausible magnetic seed field which typically has strengths of $10^{-20}\, \mathrm{G}$ is needed \citep{Durrer2013,Subramanian2016}.

In most of the gas in galaxies and on larger scales, magnetic fields are frozen into gas so we can model them with ideal magnetohydrodynamics (MHD). Magnetic fields are amplified by stretching, twisting, turning, and reconnecting them as they move with the gas, which involves magnetic resistivity that can be modelled explicitly or numerically \citep[for a recent overview about various proposed dynamo models see, e.g.][]{Rincon2019,Brandenburg2023}. 

The efficient amplification of magnetic fields leads to a strong coupling between the magnetic-field evolution and the details of the gas flows \citep{Brandenburg2005, Shukurov2006}. In the context of galaxies, this specifically applies to the inflow of gas onto galaxies, to cooling and star formation in the interstellar medium, and also to feedback processes which act on galaxy scales but can affect the whole halo.

Observations have also been claimed to show that small scale turbulent magnetic fields contribute more to the total magnetic pressure than large scale ordered magnetic fields \citep{Beck2015}, though the separation of scales is not straightforward \citep{Hollins2022}. However, it seems possible that observed variations on small scales, for example in Faraday rotation maps of the Milky Way \citep{Oppermann2015}, are caused by variations in the electron density rather than variations in the magnetic field \citep{Pakmor2018, Reissl2023}.

The full complexity of magnetic field amplification can only be addressed by high resolution cosmological simulations of galaxies. These simulations should account for the full formation history and environment of galaxies whilst, at the same time, having sufficient resolution to capture the turbulent gas flows in the interstellar medium.
Over the course of the past decade, significant progress has been made in this regard. Simulations of galaxies within a cosmological context \citep[see, e.g.][]{Guedes2011,Agertz2011,Vogelsberger2014Illustris,Schaye2005Eagle,Marinacci2014,Auriga,FIRE2,Agertz2021} are now able to produce a reasonable galaxy population which matches many properties of observed galaxies. Similarly, the inclusion of magnetic fields in these simulations has aided the understanding of the processes that amplify and shape the magnetic fields we observe in galaxies today \citep{Pakmor2014,Pakmor2017,Auriga,Rieder2017,Martin-Alvarez2018,Martin-Alvarez2022,Liu2022}.

In particular, the magnetic fields in the Auriga simulations have been shown to reproduce the strength and structure of the magnetic field in the Milky Way disc at $z=0$ reasonably well \citep{Pakmor2017,Pakmor2018}. Moreover, magnetic fields have been shown to be important for the overall evolution of galaxies at the very least in major mergers \citep{Whittingham2021,Whittingham2023}. They are also consistent with observational limits and the very recent first detection of magnetic fields in the circumgalactic medium (CGM) of the Milky Way \citep{Pakmor2020,Mannings2022,Heesen2023}.

However, similar cosmological simulations using different numerical schemes (the adaptive mesh refinement code \textsc{ramses}) and a more explicit stellar feedback model are only able to see an efficient turbulent dynamo at much higher numerical resolution and are so far not able to amplify small magnetic seed fields to saturation for Milky Way-like galaxies \citep{Martin-Alvarez2018, Martin-Alvarez2022} or dwarf galaxies \citep{Rieder2017}. Interestingly though, the strength and structure of the magnetic fields in the Auriga simulations at $z=0$ has recently been reproduced for a very similar setup with an additional subgrid model for the turbulent mean-field dynamo \citep{Liu2022} and in simulations at $z=2-1$ with much stronger initial seed fields that do not require significant amplification via a turbulent dynamo before they saturate \citep{Martin-Alvarez2021,Martin-Alvarez2023}.

Current cosmological simulations sketch a mostly coherent picture for Milky Way-like galaxies, where a high redshift turbulent dynamo amplifies a tiny magnetic seed field to roughly $\sim 10\%$ of equipartition between magnetic and turbulent energy density. Then, after the formation of a rotating gas disc, the magnetic field in this disc is further amplified to equipartition and ordered on large scales \citep{Pakmor2017}. Sufficient numerical resolution is crucial to accurately model the high redshift turbulent dynamo and to simulate its saturation before the formation of the disc \citep{Pakmor2017}. Alternatively, subgrid models for turbulent dynamos lead to very similar results \citep{Liu2022}. Note that the resolution requirements might critically depend on the numerical diffusion of the scheme used for MHD (Guillet et al. in prep).

So far cosmological zoom-in simulations that include magnetic fields have mostly focused on detailed studies of Milky Way-like galaxies \citep{Pakmor2014,Pakmor2017,Hopkins2019,Ramesh2023}. The IllustrisTNG simulations include magnetic fields at the resolution of typical cosmological box simulations focused on galaxy formation and evolution \citep[TNG100, TNG300,][]{Marinacci2018}. The highest resolution box of the IllustrisTNG simulations (TNG50, \citealt{TNG50Nelson,TNG50Pillepich}) reaches a similar mass resolution than lower resolution zoom-in simulations. Despite the lower resolution of box simulations compared to zoom-in simulations, the box simulations allow for the analysis of magnetic field amplification over a much larger range of galaxy masses and for many more galaxies.

It is important to understand the resolution requirements for modelling magnetic fields in galaxies before we put too much weight on their physical interpretation. We should confine our interpretation to properties that we know are reasonably well converged in the simulations, i.e., whose values do not change much with numerical resolution. Converged magnetic fields are also crucial for the structure of gas flows in the CGM \citep{vandeVoort2021} and for a range of other physical processes that are starting to be studied and included in new generations of galaxy simulations. 

Possibly among the most important physical processes to better model the evolution of galaxies, the modelling of cosmic ray propagation and their dynamic back reaction on gas requires reliable magnetic fields \citep{Ruszkowski_Pfrommer2023}. In particular in cosmological simulations this is especially true, where the magnetic field is not a free parameter but completely determined by gas dynamics, essentially independently from the initial seed fields if properly resolved \citep{Pakmor2014,Garaldi2021}. 

Proper resolution studies of the magnetic field evolution in cosmological simulations have so far only been done for galaxies in the mass range of the Milky Way with cosmological zoom-in simulations \citep{Pakmor2014, Pakmor2017}. Here we extend previous studies of cosmological zoom-in simulations beyond Milky Way-like galaxies by considering disc galaxies in haloes ranging in $\Mhalo$ mass from $10^{10}\,\mathrm{M_\odot}$ to $10^{13}\,\mathrm{M_\odot}$, and by simulating them at different numerical resolutions. Doing this with zoom-in simulations allows us to reach drastically higher numerical resolution for low mass galaxies than what is possible in cosmological box simulations.

We aim to understand how magnetic fields evolve over time, what minimum resolution is required to obtain converged results for our galaxy formation model, what physical processes govern the amplification and saturation of magnetic fields and set their present time structure, and how all of those change with halo mass.

The paper is structured as follows. In Section~\ref{sec:methods} we describe the simulations, simulation methods, and halo selection. In Section~\ref{sec:globalprops} we look at global properties of our disc galaxies at $z=0$. In Section~\ref{sec:amplification} we analyse the amplification of magnetic fields, its dependence on numerical resolution, and the physical properties that determine the saturation strength of magnetic fields. We then look in detail at the high redshift turbulent dynamo in galaxies over a wide range of halo masses in Section~\ref{sec:dynamo}. In Section~\ref{sec:structure}, we analyse the structure of the magnetic field in the galaxy discs at $z=0$. Finally we summarise our results and provide an outlook of future work in Section~\ref{sec:summary}.

\begin{figure*}
    \centering
	\includegraphics[width=\linewidth]{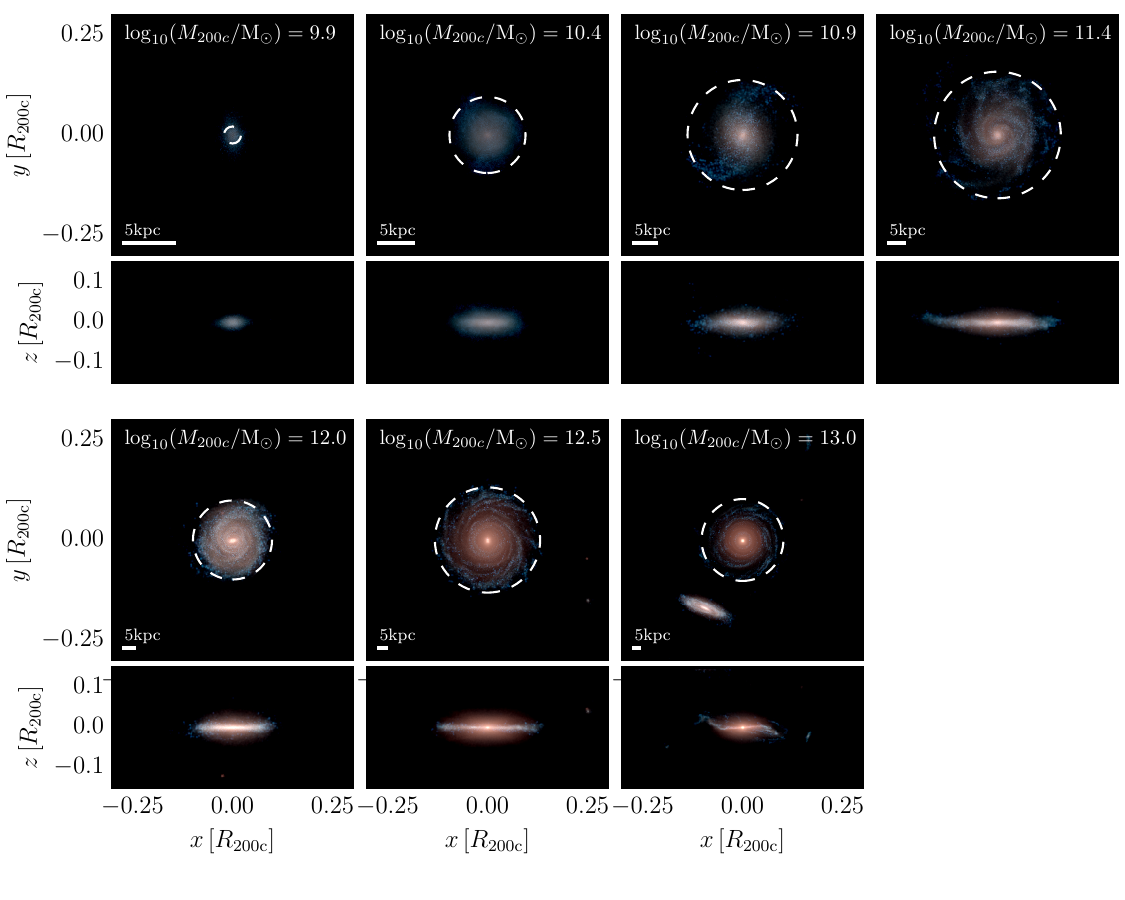}
    \caption{Face-on and edge-on stellar light projections of our galaxies at $z=0$ at the highest resolution level available for each galaxy. The $z$-axis is oriented along the eigenvector of the moment of inertia tensor of all stars within $0.1\Rhalo$ that is closest to the total angular momentum vector of those stars. We selected one galaxy per $0.5\,\mathrm{dex}$ in halo mass to illustrate the halo mass dependence and focus most of our analysis on this subsample. Dashed circles in the face-on projections show the size of the gas disc.}
    \label{fig:stellarlight}
\end{figure*}

\begin{figure*}
    \centering
	\includegraphics[width=\linewidth]{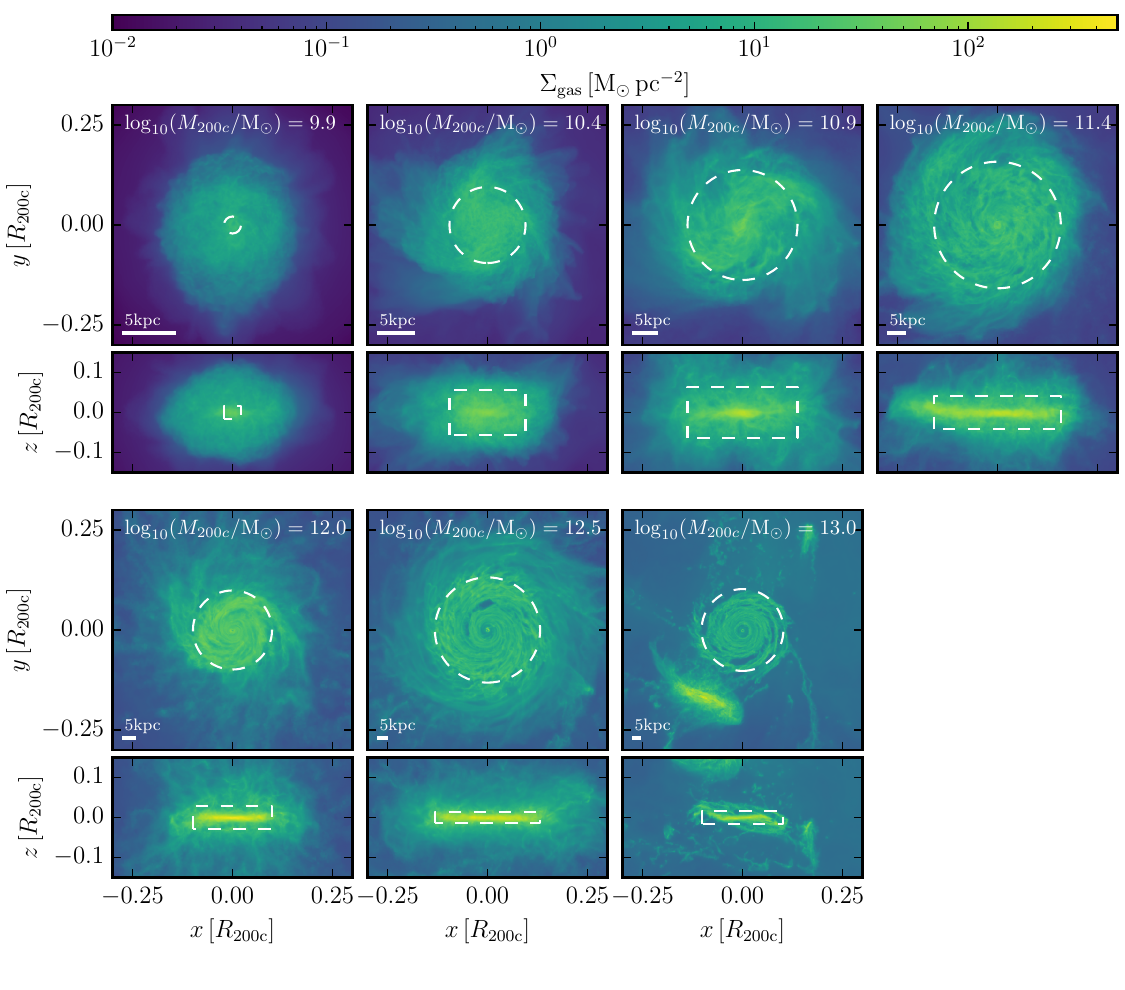}
    \caption{Face-on and edge-on gas surface density at $z=0$. Dashed circles in the face-on projections show the size of the gas disc $r_\mathrm{disc}$, defined as the circle with an average surface density of $10~\mathrm{M_\odot\,pc^{-2}}$. Dashed cylinders in the edge-on projections show radius $r_\mathrm{disc}$ and height $h_\mathrm{disc}$ of the gas disc. We discuss our definition of radius and height of the gas disc in detail in Section~\ref{sec:globalprops}. For the lowest mass halo (top left panel) we cut at $1~\mathrm{M_\odot\,pc^{-2}}$ instead.}
    \label{fig:gas}
\end{figure*}

\section{Methods and halo selection}
\label{sec:methods}

We simulate $14$ cosmological haloes with $z=0$ halo masses ($\Mhalo$) in the range $10^{10}\,\mathrm{M_\odot}$ to $10^{13}\,\mathrm{M_\odot}$ using the cosmological zoom-in technique. Here we define the halo properties $\Mhalo$ and $\Rhalo$ such that the average density in a sphere of radius $\Rhalo$ is $200$ times the critical density of the universe and $\Mhalo$ is the mass enclosed by $\Rhalo$.

We selected the haloes from the dark matter-only EAGLE box \citep{Schaye2005Eagle} and re-simulated them with a substantially increased resolution within a volume that encompasses the Lagrangian region of all dark matter particles in the halo at $z=0$ and those surrounding it out to $5\Rhalo$. To encompass the entire mass range, we selected two haloes at each $0.5~\mathrm{dex}$ in halo mass, facilitating a direct comparison with isolated galaxies analysed using a collapsing halo setup in the same range \citep{Jacob2018}. In particular, for low mass haloes we chose our haloes such that they host galaxies which are more disc dominated than is typical in this mass range. This allows us to compare more fairly and directly to the isolated galaxies which are typically disc dominated by construction.

We employ cosmological parameters from \citet{Planck2013Cosmology}, i.e. $\Omega _\rmn{m} = 0.307$, $\Omega _\rmn{b} = 0.048$, $\Omega _{\Lambda} = 0.693$ and a Hubble constant of $H_0 = 100 h$ km s$^{-1}$ Mpc$^{-1}$, where $h = 0.6777$. All quantities shown in this paper are in physical units.

We run our simulations with the \textsc{arepo} code \citep{Arepo,Pakmor2016,Weinberger2020}. \textsc{arepo} solves ideal MHD on a moving Voronoi mesh with a second order finite volume scheme \citep{Pakmor2011,Pakmor2013}. It computes self-gravity using a tree-PM scheme and couples it to MHD with a second order Leapfrog scheme \citep{Arepo,Gadget4}.
We implement additional physical processes relevant for the formation and evolution of galaxies in the \textsc{auriga} galaxy formation model \citep{Auriga}. It includes radiative cooling of hydrogen and helium, metal line cooling, heating from a spatially uniform time-evolving UV background \citep{Vogelsberger2013}, an effective model for the multiphase interstellar medium \citep{Springel2003}, a stochastic model for star formation, metal production and mass loss from star particles by stellar winds and supernovae \citep{Vogelsberger2013}, an effective model for galactic winds, as well as a model for the formation and growth of supermassive black holes and their feedback \citep{Auriga}.

The model has been widely used and tested in detail for Milky Way-like galaxies, their CGM, and for their satellites \citep[see, e.g.][]{Auriga,AurigaCGM,Simpson2018,Grand2021}. Here, we apply this model to a significantly wider range of halo masses and focus in particular on the amplification of magnetic fields. We aim to understand how the physical processes governing magnetic fields in galaxies change with halo mass and numerical resolution.

We seed magnetic fields with a spatially uniform seed field at the start of the simulations at $z=127$ with a comoving strength of $10^{-14}~\mathrm{G}$. Note that neither the exact strength of the seed field \citep{Pakmor2014} nor the detailed seeding mechanism \citep{Garaldi2021} change the properties of magnetic fields in Milky Way-like galaxies and their CGM at low redshift, provided sufficient numerical resolution is achieved. The turbulence erases all memory of the seed field at high redshift \citep{Pakmor2014,Pakmor2017} and magnetised outflows from the galaxy set the early magnetisation of the CGM \citep{Pakmor2020}. Based on establishing the conditions (in particular in terms of numerical resolution) to obtain converged magnetic fields, we augment the simulations with cosmic rays \citep{Pfrommer2016,Pakmor2016Diffusion,Buck2019} in forthcoming work (Bieri et al. in prep).

To understand the effects of numerical resolution, we ran half of the haloes (one per mass bin) additionally at different, lower resolution levels. For our simulations one resolution level lower means that all high resolution resolution elements (gas, stars, wind particles, and dark matter particles) have an eight times larger mass. The lower resolution runs use identical model parameters, in particular for star formation and feedback. We only increase all gravitational softenings by a factor of $2$ for every factor of $8$ in increase of the mass of resolution elements, as we have done in the past for the \textsc{auriga} simulations.

\begin{figure*}
    \centering
	\includegraphics[width=\linewidth]{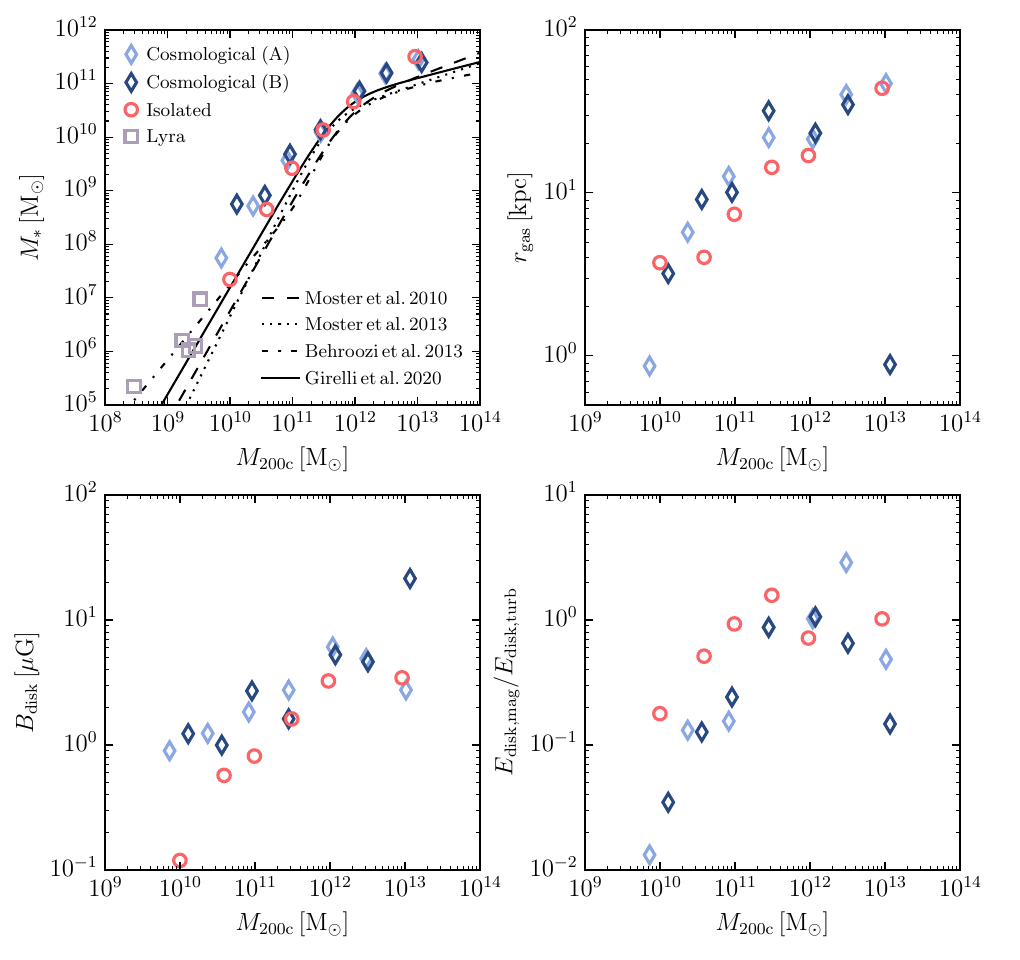}
    \caption{Stellar mass (top left panel), radius of the gas disc (top right panel), magnetic field strength in the disc (bottom left panel) and ratio between total magnetic and total turbulent energy in the disc (bottom right panel) for all cosmological galaxies at $z=0$ (blue symbols). The light blue diamonds (A) denote the subset of galaxies shown in the 7-panel figures, for example in Fig.~\ref{fig:stellarlight}. The dark blue diamonds (B) show the other $7$ halos. For comparison we also show the LYRA sample (gray squares) of cosmological dwarf galaxies \citep{Gutcke2022} and different empirical relations between stellar mass and halo mass \citep{Moster2010, Moster2013, Behroozi2013, Girelli2020} in the top left panel. Moreover, we also include the set of isolated galaxies with cosmic ray driven outflows by \citet{Jacob2018} at $3~\mathrm{Gyr}$ in all panels (red circles). The isolated galaxies agree well with the cosmological simulations over the full range of halo masses with a saturated magnetic field strength about $2$-$3$ times lower than in the cosmological simulations.}
    \label{fig:overview}
\end{figure*}

We also compare the global properties of our galaxies with isolated galaxy simulations in the collapsing halo setup \citep{Jacob2018, Pakmor2016b, Pfrommer2022}. This setup starts from a sphere of dark matter and gas that each follow an NFW mass density profile \citep{NFW1}. The gas is initially in hydrostatic equilibrium, rotating as a solid body with a dimensionless spin parameter of $\lambda=0.3$, and permeated by a uniform magnetic field with an initial strength of $10^{-10}\mathrm{G}$. From the beginning of the simulation the gas starts to cool radiatively and quickly collapses into a gas disc that forms stars and provides stellar feedback.

We focus our comparison on global properties of the galaxies and the properties of the magnetic fields. We aim to understand how well the more idealised isolated galaxy setup matches the self-consistent cosmological setup, and to get an idea to what degree we can expect the results from those isolated simulations \citep[see, e.g.][]{Pakmor2013,Pakmor2016b,Jacob2018,Pfrommer2022} to hold in cosmological galaxy simulations.

\begin{figure*}
    \centering
	\includegraphics[width=\linewidth]{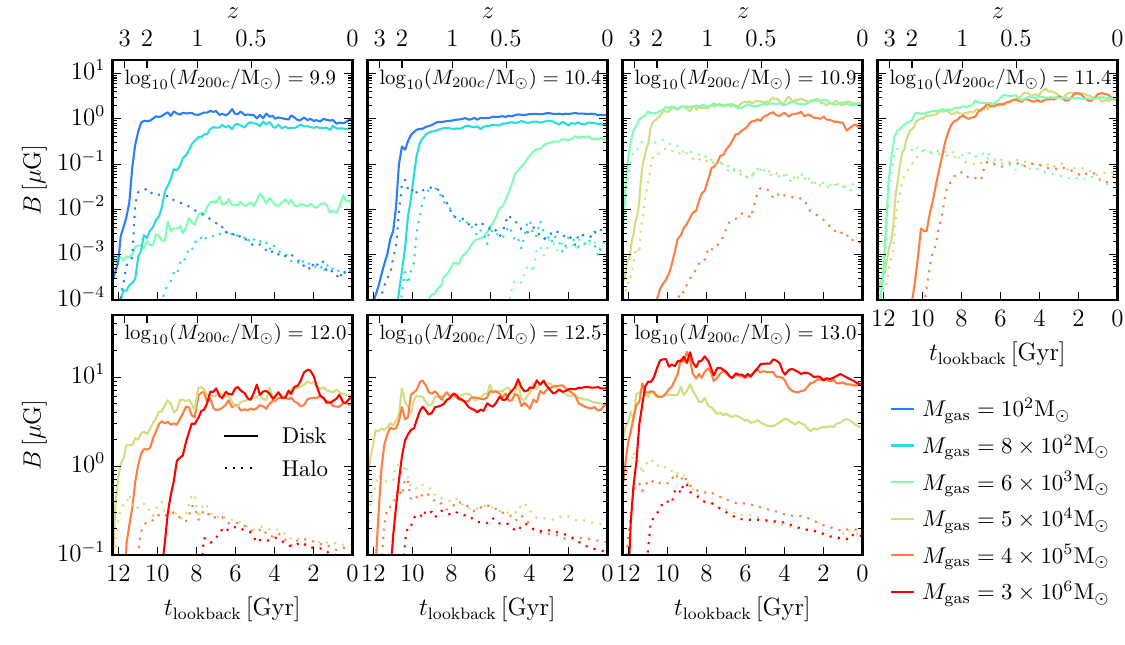}
    \caption{Magnetic field amplification over time for galaxies in haloes of different $\Mhalo$ and for different mass resolution. Solid lines show the volume-weighted root mean square magnetic field strength in the gas disc, dotted lines in the halo. The magnetic field strength in the disc is computed in a fixed physical cylindrical volume set by $r_\mathrm{disc}$ and $h_\mathrm{disc}$ at $z=0$ and rotated according to the current orientation of the stellar disc. The magnetic field strength in the halo is computed for $0.25R_{200}<r<R_{200}$ at each time.
    All galaxies show exponential amplification of the magnetic field at early times until the magnetic field saturates. The saturation strength is similar as long as a minimum required resolution is reached. This minimum resolution increases for smaller haloes. Saturation in the halo is reached at roughly the same time or only slightly later than in the galaxy.}
    \label{fig:amplification_disc}
\end{figure*}

Note, however, that this comparison is not completely fair, as the two setups have substantial differences. While the isolated and cosmological simulations use the same model for star formation and the interstellar medium, the isolated galaxies do not include an effective model for galactic winds, but instead include cosmic rays only (albeit without explicit supernova feedback), which leads to an emerging galactic wind. The isolated galaxies also do not include supermassive black holes and their associated feedback, which is problematic for the more massive galaxies. Naturally, they also lack a fully self-consistent CGM, and their stellar population is fundamentally younger. At the same time, the simulations are computationally much cheaper and can be run at a higher resolution. Here we compare to the simulations by \citet{Jacob2018} that resolve their halos with $10^6$ cells each. More recent simulations with the same setup use $10^7$ or more cells \citep{Pfrommer2022}. Additionally, a higher degree of control over the properties of the galaxy can be achieved through changes to parameters in the initial conditions.

\section{Global galaxy properties at the present time}
\label{sec:globalprops} 

We list the global properties of all cosmological haloes and their central galaxies at $z=0$ in Table~\ref{tab:galprops} for the highest resolution simulation of each halo. In the remaining parts of the paper, we will only use these highest resolution simulations, except when we explicitly study the resolution dependence.

We first focus on global properties of the galaxies and their magnetic fields and how they depend on halo mass. We show face-on and edge-on stellar light projections at $z=0$ in Figure~\ref{fig:stellarlight}. To simplify the overview and visualisation of the halo mass dependence, we display only one galaxy per mass bin.

Except for the smallest dwarf galaxy in the $10^{10}\,\mathrm{M_\odot}$ halo, all galaxies in Figure~\ref{fig:stellarlight} feature clear stellar discs at $z=0$. Even the smallest galaxy is still flattened. This is not surprising even in the lowest halo mass bins, as we selected the most disc-like galaxies (by eye), with average stellar masses (for the sample) of an original sample of $10$ galaxies per decade in halo mass to compare them with isolated disc galaxies.

We show face-on and edge-on gas surface density projections at $z=0$ in Figure~\ref{fig:gas}. Here we also orient our galaxies according to the stellar angular momentum vector for consistency. The dashed circles in the face-on projections mark the cylindrical radius that has an average surface density of $10\,\mathrm{M_\odot\,pc^{-2}}$ including all gas bound to the halo. We define this to be the disc radius. The smallest galaxy (top left panel) does not reach this threshold at all so we adopt a cut at $1\,\mathrm{M_\odot\,pc^{-2}}$ instead. The dashed rectangles in the edge-on projections show a cylinder with a radius equal to the disc radius and a height such that the averaged surface density is, again, $10\,\mathrm{M_\odot\,pc^{-2}}$. We define the disc height to be half of the height of this cylinder. We see that galaxies in haloes with $\Mhalo>10^{11.5}\,\mathrm{M_\odot}$ (top right and all bottom panels of Figures~\ref{fig:stellarlight}~and~\ref{fig:gas}) show clearly flattened discs. The galaxies in the $10^{10.5}\,\mathrm{M_\odot}$ and $10^{11}\,\mathrm{M_\odot}$ haloes show signs of discs, albeit very puffy ones.

\begin{figure*}
    \centering
	\includegraphics[width=\linewidth]{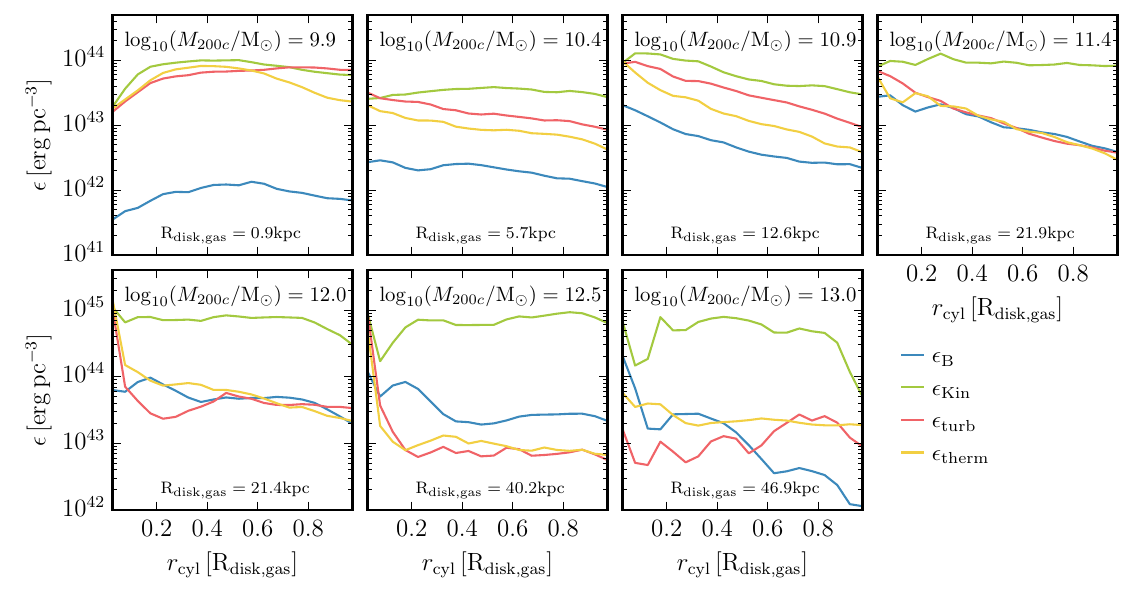}
    \caption{Radial profiles of kinetic, turbulent, magnetic, and thermal energy densities in cylindrical coordinates in the disc at $z=0$. We estimate the turbulent energy density as $1.5$ times the sum of kinetic energy densities in the radial and vertical components of the velocity field. Galaxies in haloes with $\Mhalo>10^{11.5}~\mathrm{M_\odot}$ reach equipartition between turbulent and magnetic energy density. Galaxies in smaller haloes saturate their magnetic field significantly below equipartition.}
    \label{fig:energy_disc}
\end{figure*}

We present a quantitative comparison of global properties of all cosmological galaxies with a set of isolated galaxies in Figure~\ref{fig:overview}. The cosmological galaxies are shown at $z=0$. For the stellar-mass halo-mass relation (top left panel) we also include the very high resolution cosmological dwarf galaxies of the LYRA suite \citep{Gutcke2022}, for comparison. They are run at an even higher baryonic mass resolution of $4\,\mathrm{M_\odot}$ and include cooling below $10^4\,\mathrm{K}$ and explicit thermal feedback from individual resolved supernovae instead of an effective model for the ISM. In all panels of Figure~\ref{fig:overview} we compare to the isolated galaxies from \citet{Jacob2018} at $t=3~\mathrm{Gyr}$. At later times the isolated galaxies deviate more from the cosmological galaxies as their gas reservoir depletes.

In the top left panel of Figure~\ref{fig:overview}, we see that the stellar-mass halo-mass relation is consistent between cosmological and isolated galaxies. The most massive haloes, with $\Mhalo=10^{13}\,\mathrm{M_\odot}$, have too high stellar mass compared to relations inferred from observations, likely due to shortcomings of the AGN model. At the low mass end, our galaxies are roughly consistent with the most massive galaxies from the very high resolution LYRA simulations \citep{Gutcke2022}. Note, however, that a wide range of stellar masses has been obtained in dedicated cosmological simulations of low mass haloes \citep{Onorbe2015,Maccio2017,Agertz2020}. Still it might be a bit surprising that the Auriga interstellar medium (ISM) model produces galaxies with reasonable stellar masses at this halo mass scale. We will explore the implications of this, along with a detailed comparison between effective \citep{Springel2003} and resolved \citep{Gutcke2022} ISM models at the same accessible mass scale, in future work.

In the top right panel of Figure~\ref{fig:overview} we show the sizes of gas discs at $z=0$. The sizes of the cosmological galaxies are consistent with the isolated galaxies, with the exception of one of the two galaxies in the most massive $10^{13}\,\mathrm{M_\odot}$ haloes  (1e13\_h7) that has essentially no gas left in its central galaxy at $z=0$, likely due to feedback from active galactic nuclei.

The bottom left panel of Figure~\ref{fig:overview} shows the volume averaged root mean square magnetic field strength in the disc at $z=0$. It is qualitatively similar between cosmological and isolated galaxies, but a factor of $2$-$3$ higher in the cosmological galaxies in comparison to the isolated galaxies. For both sets there is only a weak trend of magnetic field strength in the disc with halo mass. Only the smallest galaxy of the isolated sample has a significantly lower magnetic field strength.

Finally, the bottom right panel of Figure~\ref{fig:overview} shows the ratio of the total magnetic energy to the total turbulent energy in the disc at $z=0$. Here, we estimate the turbulent energy density as $1.5$ times the sum of the kinetic energy densities in the radial and vertical components of the velocity field, assuming that these components are dominantly turbulent, and the azimuthal component of the velocity field has a turbulent contribution of equal strength. We focus on the comparison between total magnetic energy and total turbulent energy, because we expect the turbulent dynamo to saturate when the magnetic energy density reaches several $10\%$ of the turbulent energy density, though it might be as high as unity under certain conditions \citep{Federrath2016, Kriel2023}. Similarly, for most large scale dynamos the small scale velocity field likely sets the saturation strength of the total magnetic field \citep{Rincon2019}. Moreover, equipartition between total magnetic energy density and turbulent energy density is consistent with observations of nearby galaxies \citep[see, e.g.][]{Beck2015}.

In cosmological simulations of Milky Way-like galaxies we have seen saturation of the turbulent dynamo at high redshift when the magnetic energy density reaches a few $10\%$ of the turbulent energy density. We then observe further amplification to equipartition between turbulent and magnetic energy density after the formation of a rotating gas disc \citep{Pakmor2017}. Except for the galaxies in the lowest mass haloes $\Mhalo\leq 10^{10}\,\mathrm{M_\odot}$, all cosmological galaxies as well as all isolated galaxies reach at least $10\%$ of equipartition between magnetic and turbulent energy. Galaxies in haloes with $\Mhalo\geq 10^{11.5}\,\mathrm{M_\odot}$ are consistent with equipartition, except for the 1e13\_h7 that has no dense starforming gas left at $z=0$.

\begin{figure*}
    \centering
	\includegraphics[width=\linewidth]{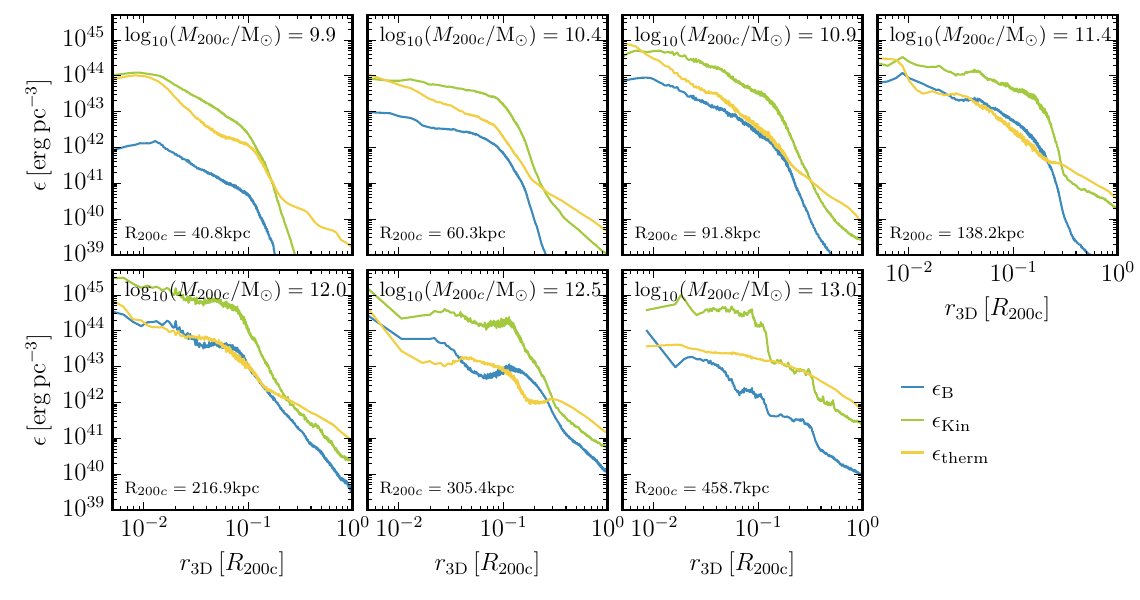}
    \caption{Radial profiles of kinetic, magnetic, and thermal energy densities in spherical coordinates in the whole halo at $z=0$, excluding satellite galaxies. In haloes more massive than $\Mhalo>10^{11}~\mathrm{M_\odot}$ the magnetic energy density saturates at around $10\%$ of the kinetic energy density in the CGM. For smaller haloes, the magnetic field in the CGM is barely amplified at all. At the smallest radii, where the profile measures the energies in the central galaxies, the magnetic energy density reaches equipartition with the thermal energy density for haloes more massive than $\Mhalo>10^{11}~\mathrm{M_\odot}$.}
    \label{fig:energy_halo}
\end{figure*}

The isolated galaxies show comparable levels of saturation of the magnetic energy density for $\Mhalo\geq 10^{11.5}\,\mathrm{M_\odot}$. For galaxies in less massive haloes the saturation level is systematically higher in isolated galaxies. Together with the bottom left panel of Figure~\ref{fig:overview}, which shows the magnetic field strength, this paints a consistent picture in which the turbulent energy density is significantly lower in isolated galaxies. This is likely due to two combined reasons. Turbulent driving by accretion/shear against the CGM is reduced, as the isolated galaxies have been depleted of gas. Moreover, isolated galaxies have weaker outflows because of the absence of the effective galactic wind model, which plays an important role in the cosmological galaxies. The general good agreement in many properties between cosmological and isolated galaxies is quite surprising given the differences in feedback models, but clearly demonstrates the usefulness of the collapsing halo setup to study galaxy formation and evolution in comparison to setups that do not follow the formation of galactic discs self-consistently, but require additional assumptions for example on the properties the magnetic fields (like isolated discs or tallbox setups).

\section{Amplification and saturation of magnetic fields}
\label{sec:amplification}

For Milky Way-like galaxies we have found in previous work that the magnetic field amplification proceeds in two steps. Once the galaxy has grown sufficiently large, a turbulent dynamo amplifies the magnetic field at high redshift. It saturates when the magnetic energy density reaches a few $10\%$ of the turbulent energy density. For sufficiently high resolution this happens long before the galaxy forms a disc. Later, after the formation of a disc, the magnetic field is ordered and further amplified to equipartition where it remains until $z=0$ \citep{Pakmor2014,Pakmor2017}. The CGM is magnetised slightly later than the central galaxy, primarily by magnetised outflows from the central galaxy, until at low redshift an in-situ dynamo in the halo contributes as well \citep{Pakmor2020}. We now aim to understand to what degree this picture holds in less massive and more massive haloes, and whether we need to adapt it.

\begin{figure*}
    \centering
	\includegraphics[width=\linewidth]{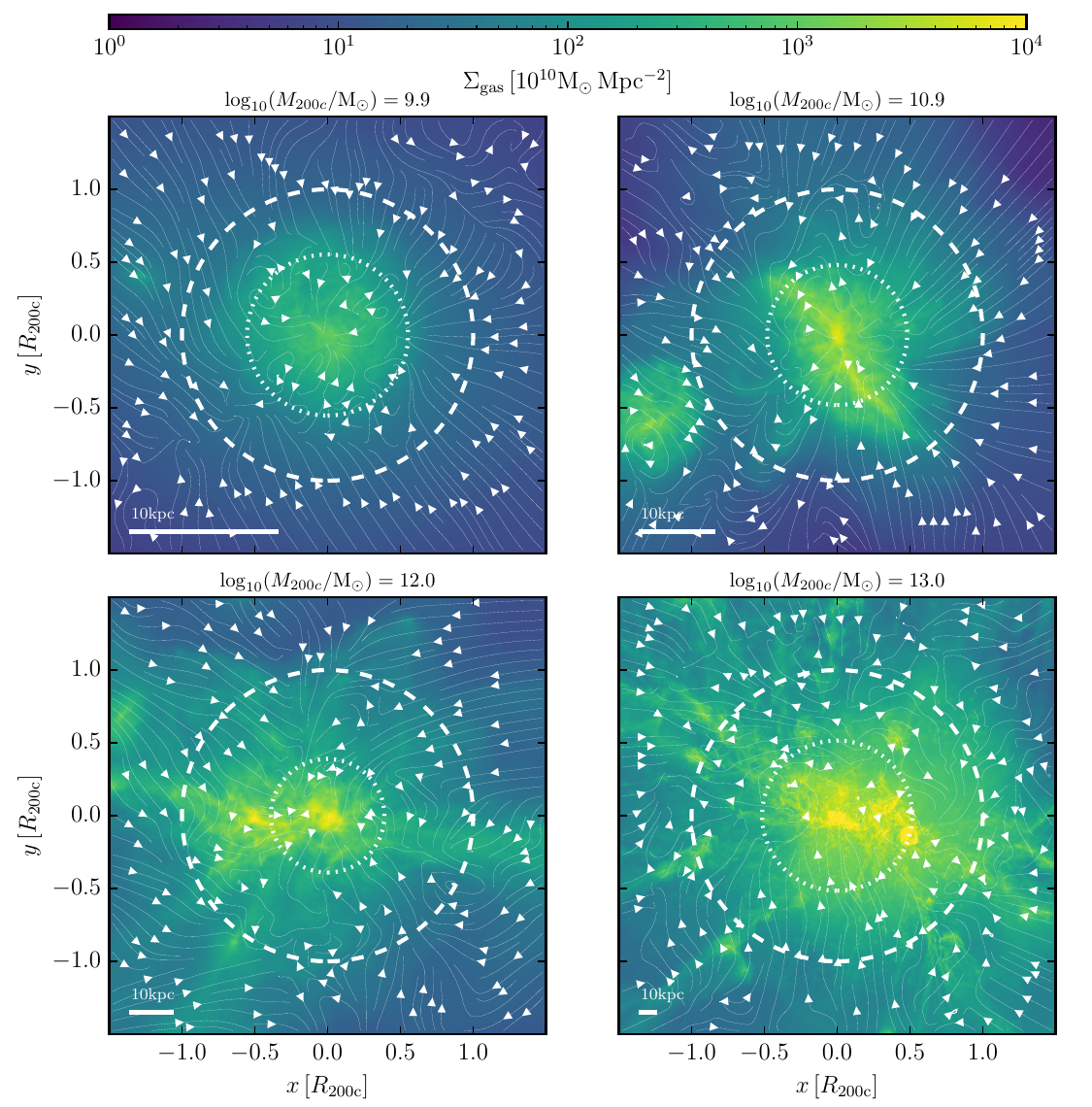}
    \caption{Gas surface density projections of four haloes, increasing in mass from $10^{10}\,\mathrm{M_\odot}$ to $10^{13}\,\mathrm{M_\odot}$ at $z=3$ with velocity streamlines. The dashed circles show $R_{200}$ at this time. The dotted circles give the radius where wind particles typically recouple at this time. All haloes feature large scale coherent inflows onto the halo that transition to a turbulent flow close to the recoupling radius where inflows and outflows meet.}
    \label{fig:gasvel}
\end{figure*}

In Figure~\ref{fig:amplification_disc} we show the time evolution of the magnetic field strength in the disc (solid lines) and the halo (dotted lines) for one halo per mass bin, and for three different numerical resolution levels per halo. We compute the magnetic field of the disc in the physical volume of the disc cylinder at $z=0$, as defined in Section~\ref{sec:globalprops}, even if there is no disc yet. We orient the cylinder according to the stars at the current time. We compute the magnetic field in the halo in a volume given by $0.25\,\Rhalo<r<\Rhalo$, where we use the value of $\Rhalo$ at that time, meaning that the physical volume we consider changes with time. Note that for different resolution levels the disc sizes change by less than a factor of $2$ and typically only by $20\%-30\%$, and that $\Rhalo$ is essentially identical. 

Figure~\ref{fig:amplification_disc} shows a clear exponential amplification of the magnetic field in galaxies of all halo masses at high redshift, provided that the simulations are run with sufficient numerical resolution. In this case, the magnetic field strength saturates at essentially the same strength, independent of resolution. This resolution requirement, however, depends significantly on the halo mass.

The converged saturated magnetic field strength in all galaxies is several orders of magnitude larger than expected from purely adiabatic changes of the initial seed field. For purely adiabatic changes, i.e. the magnetic field strength scales with $\rho^{2/3}$ (see also Equation~\ref{eq:b_ad} for a more detailed discussion), we would roughly expect a magnetic field strength of $10^{-4}\mu \mathrm{G}$ at densities of $1\,\mathrm{cm^{-3}}$ that are typical for the ISM. The magnetic field strengths we see in our galaxies (e.g. in Figure~\ref{fig:amplification_disc}) are at least of order $\mu \mathrm{G}$, which requires an efficient dynamo to operate.

For the lowest mass haloes, $\Mhalo\leq 10^{10.5}\,\mathrm{M_\odot}$, the simulation with a mass resolution of $M_\mathrm{gas}=6\times 10^{3}\,\rmn{M}_\odot$ is insufficient to properly saturate the magnetic field strength, the simulation with $M_\mathrm{gas}=8\times 10^{2}\,\rmn{M}_\odot$ seems to reach the physical saturation level only at low redshift (after $z=1$), and only the highest resolution simulation with a mass resolution of $M_\mathrm{gas}= 10^{2}\,\rmn{M}_\odot$ has a saturated magnetic field strength at high redshift.

For haloes with masses $\Mhalo=10^{11}\,\mathrm{M_\odot}$ and $\Mhalo=10^{11.5}\,\mathrm{M_\odot}$ we find that we need a resolution of $5\times 10^{4}\,\mathrm{M_\odot}$ to saturate the magnetic field strength at sufficiently high redshift, well before a gas disc forms. For Milky Way-like galaxies with a halo mass close to $\Mhalo=10^{12}\,\mathrm{M_\odot}$ a resolution of $4\times 10^{5}\,\mathrm{M_\odot}$ is sufficient, and for more massive haloes even at a mass resolution of $3\times 10^{6}\,\mathrm{M_\odot}$ the magnetic field is quickly amplified and saturates at high redshift. For the most massive halo, the magnetic field strength is lower at the highest resolution because the disc contains significantly less gas.

\begin{figure*}
    \centering
	\includegraphics[width=\linewidth]{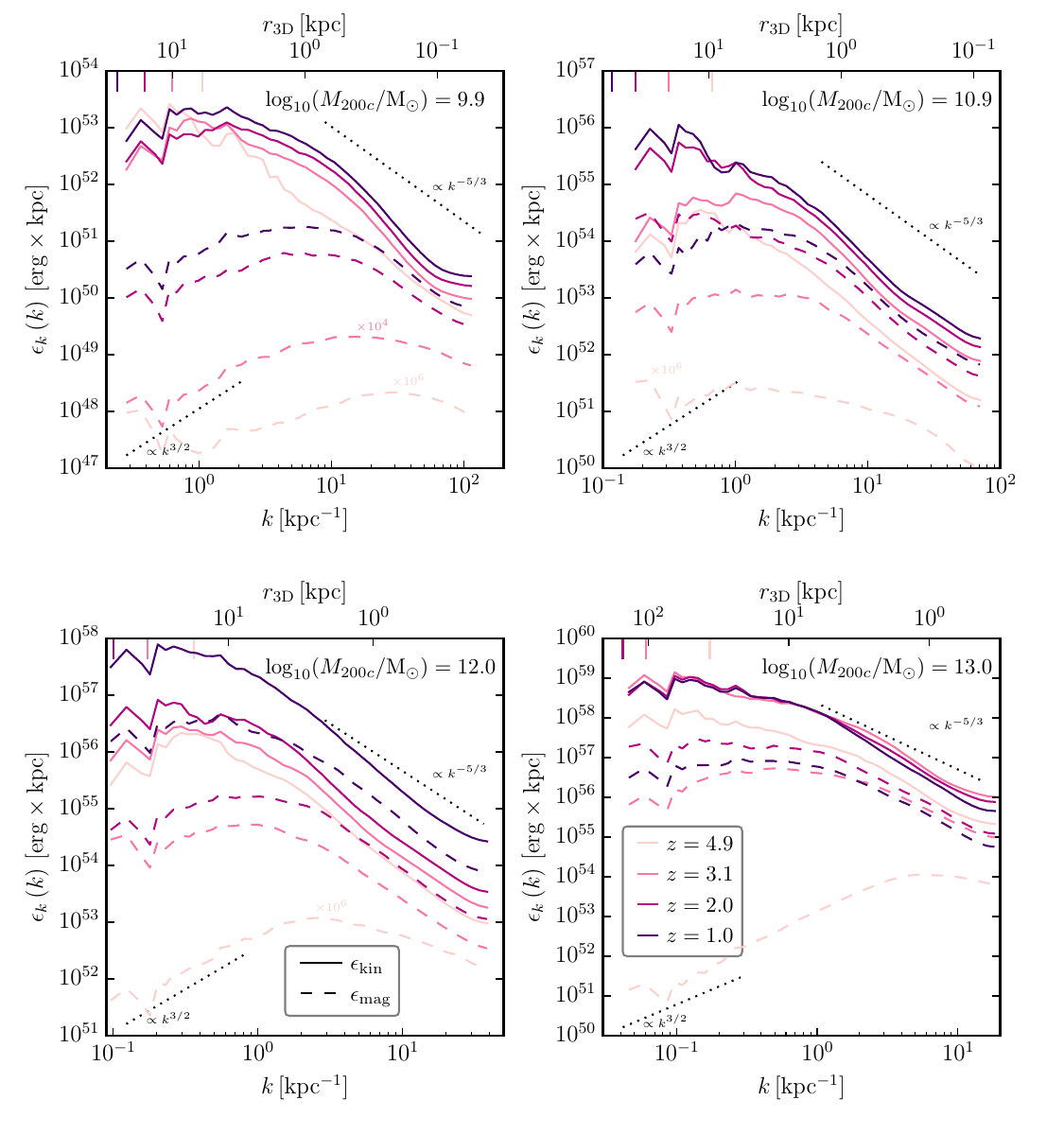}
    \caption{Kinetic (solid lines) and magnetic (dashed lines) power spectra of four haloes from $z=5$ to $z=1$. We compute the power spectra in a sphere of fixed physical radius set by $R_{200}$ at $z=1$. Dotted lines show the expected slopes of a Kazantsev spectrum \citep[$\propto k^{3/2}$,][]{Kazantsev1985} and a Kolmogorov spectrum \citep[$\propto k^{-5/3}$,][]{Kolmogorov1941}. Vertical lines on the top show $\Rhalo$ at each time. The power spectra of all galaxies are consistent with a turbulent dynamo that amplifies the magnetic field. It saturates first on small scales, then over time also on larger and larger scales.}
    \label{fig:powerspectra}
\end{figure*}

\begin{figure*}
    \centering
	\includegraphics[width=\linewidth]{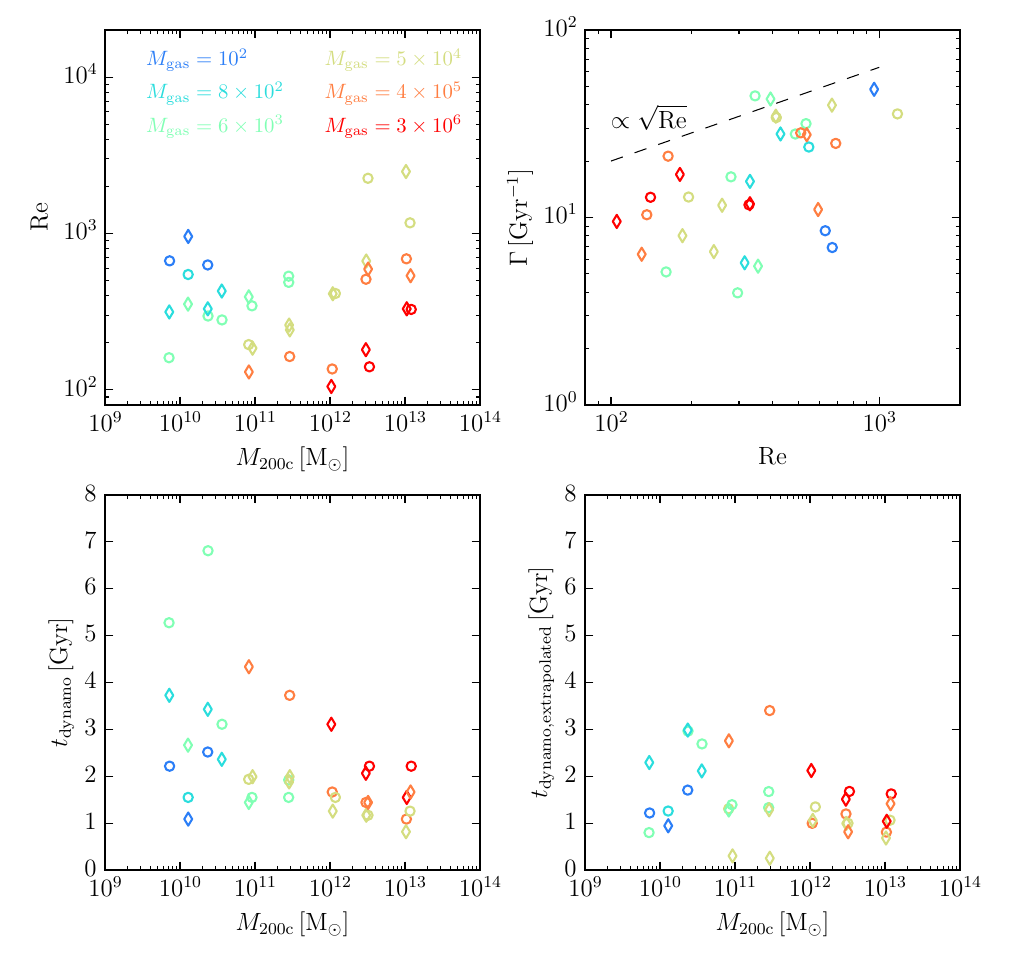}
    \caption{Quantification of the turbulent dynamo phase. The top panels give an estimate of the Reynolds number (top left panel, see Equation~\eqref{eq:reynolds} for the estimate we use) and the correlation between Reynolds number and amplification rate of the magnetic energy density (top right panel). The circles and diamonds correspond to different simulations from the Cosmological (A) and Cosmological (B) sets, respectively. The colors denote the baryonic mass resolution of the simulations, and $\Mhalo$ is the mass of the halo at $z=0$. We estimate the time when the magnetic energy in $10\%$ of the virial radius first exceeds $1000$ times the energy expected for purely adiabatic changes of the magnetic field strength. We measure the amplification rate at a similar time, for details see the text. The bottom panels show the age of the universe at this time when the dynamo kicks in $t_\mathrm{dynamo}$ (bottom left panel) and the time when the amplification started, extrapolating back from the $t_\mathrm{dynamo}$ under the assumption that the amplification rate is constant (bottom right panel). The amplification rate scales as expected for a turbulent dynamo. Moreover, we argue that the start of the exponential amplification of the magnetic field is set by the evolution of the halo, rather than reaching a critical Reynolds number.}
    \label{fig:bfld_amplification_overview}
\end{figure*}

Note that these quantitative resolution requirements almost certainly depend on the details of the numerical scheme, in particular its numerical viscosity and resistivity. The resolution requirements will also depend on the galaxy formation and feedback model employed, and are directly applicable only for simulations ran with the Auriga model \citep{Auriga}. Nevertheless, they can at least give us an idea to what degree we can trust the magnetic field properties in the large cosmological box simulations of the IllustrisTNG project \citep{Marinacci2018} that employ a galaxy formation model \citep{TNGMethodWeinberger,TNGMethodPillepich} which is, in many aspects, similar to the Auriga model. Directly adopting our resolution requirements, we can infer that the magnetic field of galaxies in TNG100, that used a mass resolution of $10^{6}\,\mathrm{M_\odot}$, is reliable down to halo masses of $\Mhalo\gtrsim 10^{12}\,\mathrm{M_\odot}$, and in galaxies in TNG50 with its mass resolution of $8\times 10^{4}\,\mathrm{M_\odot}$ down to halo masses of $\Mhalo\gtrsim 10^{11}\,\mathrm{M_\odot}$.

\begin{figure*}
    \centering
	\includegraphics[width=\linewidth]{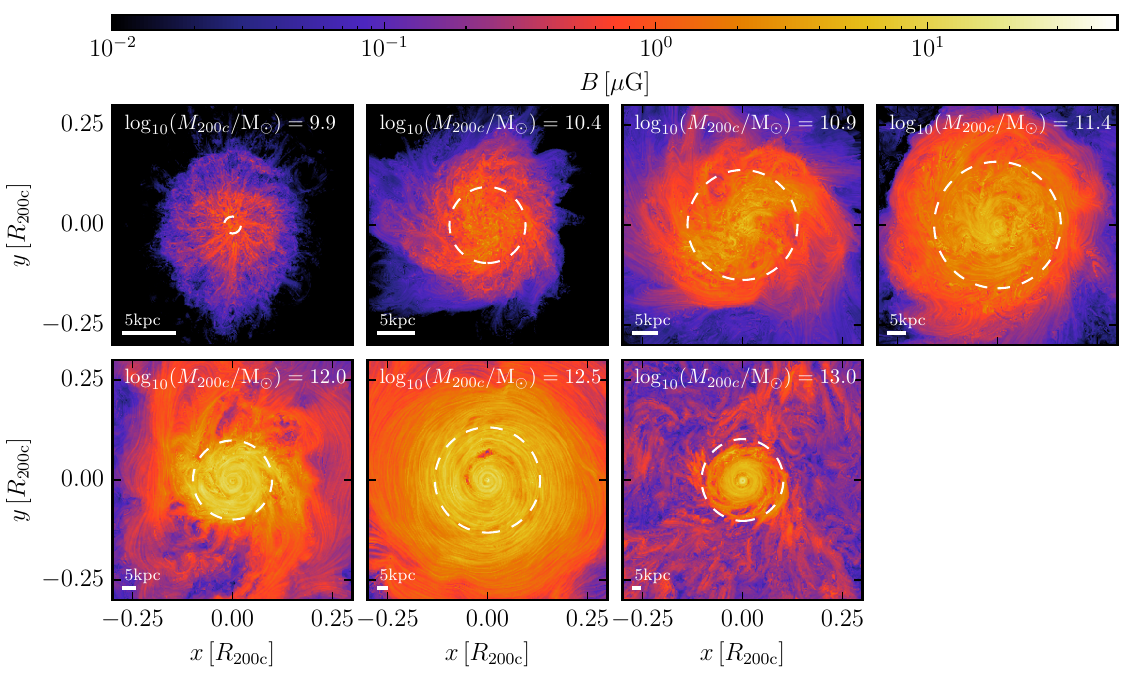}
    \caption{Magnetic field strength at $z=0$ in thin face-on projections. The relief shows the orientation of the magnetic field strength using the line integral convolution method \citep{LIC}. The ordering of the magnetic field increases with halo mass, from being essentially unordered to showing predominantly azimuthal magnetic fields.}
    \label{fig:bfld}
\end{figure*}

We next look for the physical cause of the saturation strength of the magnetic field. In Figure~\ref{fig:energy_disc} we compare the radial profiles of different energy densities at $z=0$ in cylindrical shells with the height of the gas disc of the highest resolution simulation of each galaxy. The saturation strength is consistent with equipartition between magnetic energy density and thermal or turbulent energy density in galaxies of haloes with $\Mhalo\geq 10^{11.5}\,\mathrm{M_\odot}$.
For smaller haloes a dynamo is still clearly present, but the magnetic energy density saturates below equipartition with the thermal or turbulent energy density. This gap widens for progressively lower halo masses.

Interestingly, in the galaxies that reach equipartition the kinetic energy in ordered rotation clearly dominates over turbulent, thermal, and magnetic energy. For galaxies in less massive haloes ordered rotation becomes less and less important, and at the same time the ratio between magnetic energy density and turbulent and thermal energy density decreases. For the galaxy in the smallest $10^{10}\,\mathrm{M_\odot}$ halo the kinetic energy is dominated by turbulent energy, and ordered rotation is mostly absent for the gas. For this galaxy the magnetic field saturates at only a few percent of equipartition. Importantly, the saturation strength seems to be set by physical properties of the galaxy for all galaxies, because it is converged with resolution (see Figure~\ref{fig:amplification_disc}).

The magnetic field strength in the halo is shown in Figure~\ref{fig:amplification_disc} by the dotted lines. The picture that emerges is very similar to the evolution of the magnetic field strength in the central galaxy, but slightly delayed. The resolution requirements to obtain a converged saturated magnetic field in the halo are essentially the same as for the magnetic field strength in the central galaxy. The magnetic field in the halo always starts to be amplified and saturates after the magnetic field in the galaxy. This indicates that magnetised outflows from the central galaxy whose magnetic field has been amplified already, contribute significantly at all halo masses, consistent with earlier results for the halo of Milky Way-like galaxies \citep{Pakmor2020}. The apparent drop of the magnetic field strength in the halo in Figure~\ref{fig:amplification_disc} after saturation is caused by the growth of the halo, so that the magnetic field is averaged over a larger volume that extends to lower density gas.

We show radial profiles of different energy densities in the whole halo at $z=0$ in Figure~\ref{fig:energy_halo}. We compute the energy densities in spherical shells in the rest frame of the halo and include only cells that are associated with the main subhalo, i.e. excluding cells bound to satellite galaxies. Haloes with $\Mhalo\geq 10^{11}\,\mathrm{M_\odot}$ have significantly amplified magnetic fields in the CGM out to at least the virial radius. The saturation level of the magnetic energy in the CGM ($0.2\Rhalo \lesssim r_\mathrm{3D} \lesssim \Rhalo$) is roughly $10\%$ of equipartition with the (fluctuating) kinetic energy. The most massive halo with $\Mhalo=10^{13}\,\mathrm{M_\odot}$ seems to saturate at a slightly smaller fraction of equipartition. Haloes with $\Mhalo\leq 10^{10.5}\,\mathrm{M_\odot}$ essentially do not amplify the magnetic field in the halo, and the magnetic energy density profile drops steeply at about $0.2\,\Rhalo$. Likely this is a result of the combination of very weak galactic outflows and weaker inflows that lead to less turbulent haloes at low redshift at this mass scale in the Auriga model. Many of the profiles display a feature around $r_\mathrm{3D} \sim 0.2\Rhalo$ that is connected to the size of the gas disc. For smaller radii the kinetic energy density is dominated by rotation. For larger radii instead the kinetic energy density corresponds to the turbulent energy density in the absence of large-scale ordered rotation. Moreover, the average gas density drops significantly at the edge of the gas disc.

From Figure~\ref{fig:energy_halo} we can also infer characteristic values for the plasma beta $\beta = P_\mathrm{therm}/P_{B}$. We find typical values for $\beta$ between $10$ and $100$ in the CGM at all radii for haloes with masses $\Mhalo\geq 10^{11}\,\mathrm{M_\odot}$ and much larger values for the smaller haloes.

\section{The omnipresent high redshift turbulent dynamo}
\label{sec:dynamo}

We showed in Figure~\ref{fig:amplification_disc} that exponential amplification of the magnetic fields begins (given sufficient numerical resolution) at high redshift, at lookback times of $t_\mathrm{lookback}>10~\mathrm{Gyr}$. We also discussed already in Section~\ref{sec:amplification} that the magnetic field becomes orders of magnitudes stronger than expected from adiabatic compression alone.

The obvious candidate for a physical mechanism that quickly amplifies the magnetic field at early times is a small-scale dynamo, which is the main source of the amplification process in Milky Way-like galaxies \citep{Pakmor2017}. For a turbulent dynamo a turbulent velocity field is essential, leading to the exponential amplification of the magnetic field. This amplification continues until saturation occurs, initially on smaller scales, and subsequently on larger ones, when the magnetic energy density reaches $\sim 10\%$ of the turbulent energy density on a given scale \citep{Federrath2016}.

To get a qualitative idea about the structure of the velocity field at high redshift we turn to Figure~\ref{fig:gasvel}, which shows the surface density and velocity field at $z=3$ for four haloes in the mass range $10^{10}\,\mathrm{M_\odot}$ to $10^{13}\,\mathrm{M_\odot}$ projected along the $z$-axis of the simulation box. The picture that emerges holds for the other halos as well. All haloes show a very similar picture of large-scale coherent inflows feeding the galaxy with gas. The in-flowing gas moves in from different directions, interacts with other inflowing gas as well as outflowing gas from the central galaxy and produces a highly chaotic, turbulent velocity field at smaller radii. The transition to a turbulent flow happens roughly where the wind particles re-couple and deposit their mass, momentum, and thermal energy into a cell, which is set by a threshold density of $0.05\,\mathrm{cm^{-3}}$. At $z=5$ this density threshold translates to a radius of $\sim 0.75\,\Rhalo$, and at $z=3$ to about $0.5\,\Rhalo$. At $z=0$ they typically recouple within a few kpc of the disc.

There are quantitative differences between haloes of different masses. The turbulent region extends to larger radii in more massive haloes, and turbulence is already stronger. 

Having established qualitatively that the velocity field at the centre of the haloes is turbulent at high redshift, we now quantify the structure of the velocity and magnetic field in Figure~\ref{fig:powerspectra}. This figure shows power spectra of the kinetic energy density (solid lines) and magnetic energy density (dashed lines) for the same haloes as in Figure~\ref{fig:gasvel} for four different redshifts from $z=5$ to $z=1$. To obtain the kinetic power spectra we first map each component of $\sqrt{\rho/2}\,\bs{\varv}$ to a $1024^{3}$ uniform Cartesian mesh. We chose the size of the mesh as twice $\Rhalo$ at $z=1$ and keep the same physical size for all power spectra of one halo. We use zero padding to compensate for the periodicity of the Fourier transformation. We then transform each component into Fourier space where we compute $\epsilon_{k}$ as the one-dimensional kinetic energy density averaged in shells in $k$ space. We use the same procedure to obtain the magnetic power spectra from mapping $\bs{B}/\sqrt{8\pi}$. The kinetic power spectra are consistent with subsonic turbulence \citep{Kolmogorov1941} with an injection scale only slightly smaller than the size of the halo at the higher redshifts (e.g. $z=3$). The energy injected on large scales cascades down to the grid scale which assumes the smallest physical dimensions in the ISM for all haloes. The haloes differ quantitatively in how quickly the kinetic energy density increases with time, and how the peak of the kinetic power spectrum shifts with time.

At $z=1$ the central galaxies in the more massive haloes have formed gas discs. The initial strong gas accretion in all haloes has calmed down and the haloes have become predominantly hydrostatic. Feedback driven outflows from the central galaxy have become stronger as the star formation rate has substantially increased, but those are not the exclusive drivers of turbulence in the halo. Turbulence in the central galaxy is probably dominantly sourced by star formation, and the interaction between the disc and the halo \citep{Bieri2022,Pfrommer2022} for the galaxies with a gas disc.

The shape of the magnetic power spectra are consistent with expectations of a turbulent dynamo. The magnetic field strength increases exponentially with time. It grows fastest on the smallest scales and then saturates there when the magnetic energy density reaches $\sim 10\%$ of the kinetic energy density on small scales. This process happens on larger scales as well, but takes a longer time. For the lowest mass haloes the largest scale on which the magnetic field eventually reaches equipartition is small compared to the size of the halo and at a smaller fraction of the injection scale. Therefore the total magnetic energy in the lower mass haloes saturates at a smaller fraction of the total kinetic energy than in massive haloes.

The slope of the magnetic power spectrum on large scales is consistent with $\propto k^{3/2}$ which is as expected for the kinematic amplification phase of the turbulent dynamo \citep{Kazantsev1985}. Over time, the saturation scale grows to larger and larger scales until it reaches scales just below the injection scale. Note that, at $z=1$, the kinetic power spectrum shown in Figure~\ref{fig:powerspectra} includes kinetic energy in ordered differential rotation in the gas disc if present, i.e. in the more massive haloes.

Having established the presence of a turbulent dynamo, we can try to better understand the resolution effects seen in Figure~\ref{fig:amplification_disc}. We see two main effects. At higher resolution the amplification of the magnetic field strength proceeds faster and it seems to start earlier.

From idealised simulations of subsonic turbulence we expect that a minimum numerical Reynolds number of $\mathrm{Re}\gtrsim 100$ is needed to amplify the field \citep{Schober2012}. For smaller Reynolds numbers dissipation is faster than amplification and the turbulent dynamo is completely suppressed. Moreover, we expect the amplification rate of the magnetic field strength in the kinetic regime to roughly scale with $\Gamma \propto \sqrt{\mathrm{Re}}$ \citep{Schober2012,Pfrommer2022}.

We now confront both expectations with the magnetic field evolution in our galaxies and at different numerical resolution. Note, however, that the situation in our simulations is much more complicated and the comparison should be interpreted with appropriate care. In particular, there is more than one process driving turbulence, at least gravitational infall and feedback driven outflows from the galaxy, potentially driving on different scales. Moreover, the gas density as well as the numerical resolution in the haloes change significantly with distance to the center of the haloes, and the turbulent flow is multiphase and potentially coupled via magnetic fields.

To quantify the magnetic field amplification in our simulations we first find the earliest snapshot in which the magnetic energy within $0.1\Rhalo$ is larger than $1000$ times the value it had from purely adiabatic compression. For this estimate we compute the adiabatically expected magnetic field strength of all cells in $0.1\Rhalo$ as 
\begin{equation}
    B_\mathrm{ad} = B_0 \left( \frac{\rho}{\rho_0} \right)^{2/3},
    \label{eq:b_ad}
\end{equation}
where $B_0$ is the physical magnetic field strength in the initial conditions and $\rho_0$ the physical gas density at this time. When this condition is met, consequently the magnetic field strength has been amplified by more than $\sqrt{1000}$, i.e. more than an order of magnitude already by a dynamo. The magnetic energy density is still orders of magnitude smaller than the turbulent energy density, so we are safely in the kinetic regime where the back reaction of the magnetic field on the gas dynamics is irrelevant. We call this time $t_\mathrm{dynamo}$. We estimate the numerical Reynolds number at this time as
\begin{equation}
    \mathrm{Re}\approx \frac{3 \mathcal{L}}{d_\mathrm{cell}} \frac{\mathcal{V}}{\varv_\rmn{th}},
    \label{eq:reynolds}
\end{equation}
where $\mathcal{L}$ is the length scale of turbulent injection, $d_\mathrm{cell}$ is the typical diameter of the smallest cells in the center of the galaxy, and we assume $\mathcal{V}\approx\varv_\rmn{th}$ \citep{Pfrommer2022}. For $\mathcal{L}$ we take the radius at which the wind particles recouple (see also Figure~\ref{fig:gasvel}) and for $d_\mathrm{cell}$ we use the average diameter of the $100$ smallest cells in the halo as an estimate of the smallest eddies we resolve. Using the average of the $10$ or $1000$ smallest cells gives essentially identical results.

To estimate the amplification rate of the magnetic field we combine the snapshot at $t_\mathrm{dynamo}$, the last snapshot before it, and all later snapshots until the magnetic energy density in the same physical volume that we used to find $t_\mathrm{dynamo}$ has exceeded $10^4$ times the adiabatically expected magnetic energy density. We fit an exponential growth rate to the average magnetic field strength in the same volume of all those snapshots as
\begin{equation}
    \epsilon_B(t) = \epsilon_{B,0}\ \rmn{e}^{\Gamma t},
\end{equation}
where $\epsilon_B(t)$ is the magnetic energy density at time $t$, $\epsilon_{B,0}$ is the magnetic energy density expected from purely adiabatic changes of the magnetic field, and $\Gamma$ is the amplification rate of the magnetic energy density.

\begin{figure*}
    \centering
	\includegraphics[width=\linewidth]{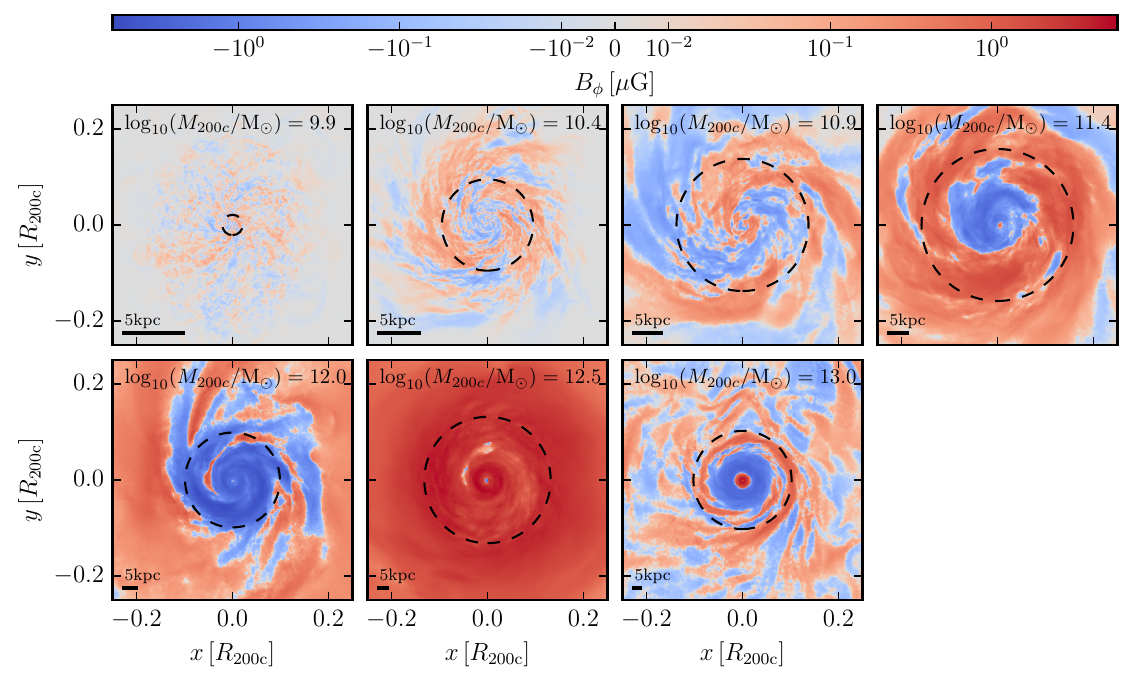}
    \includegraphics[width=\linewidth]{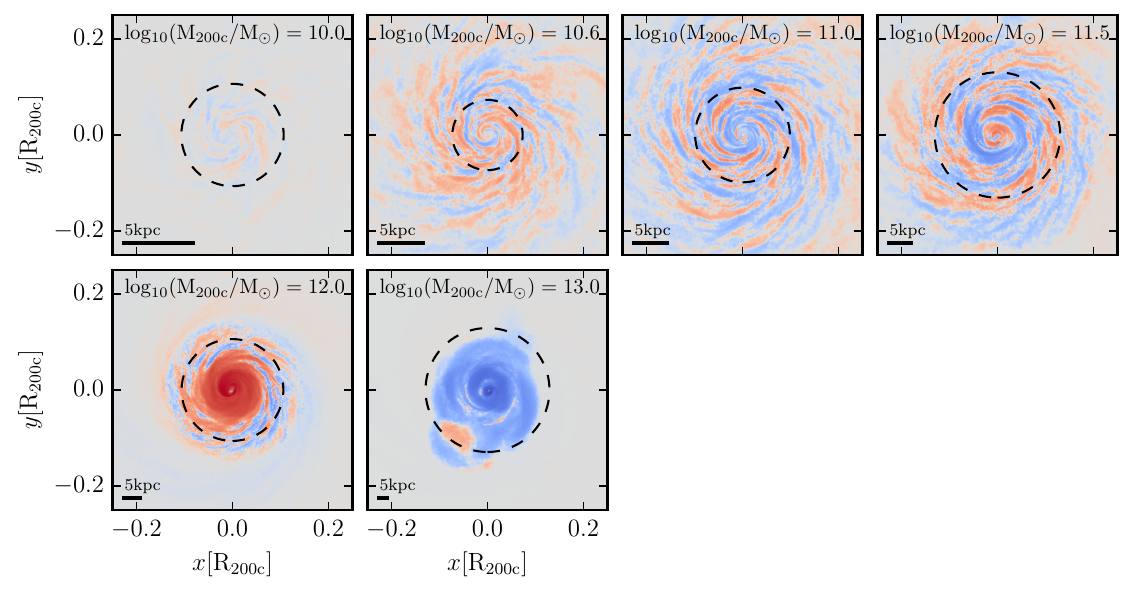}
    \caption{Azimuthal component of the magnetic field of the cosmological galaxies at $z=0$ (top panels) and of the isolated galaxies at $3\,\mathrm{Gyr}$ (bottom panels) in thin face-on projections. The black dashed circle shows the size of the gas disc. The azimuthal magnetic field becomes more ordered in more massive halos in both sets of simulations.}
    \label{fig:bphi}
\end{figure*}

We show the amplification properties for all our galaxies at all available resolution levels in Figure~\ref{fig:bfld_amplification_overview}. They all indicate a turbulent dynamo with Reynolds numbers $\mathrm{Re}\geq 100$. The Reynolds number increases with increasing resolution for the same halo, and increases as well with halo mass at fixed resolution. The increase with resolution is easily explained as the size of the smallest cells decreases with resolution but the injection scale stays roughly the same. Note though that the Reynolds numbers are estimated at different times (as shown in the bottom left panel of Figure~\ref{fig:bfld_amplification_overview}), so the objects are not directly comparable. The global properties usually only change slowly, however, on timescales of Gyrs.

We show the growth rate of the magnetic energy in the top right panel of Figure~\ref{fig:bfld_amplification_overview}. We see that the growth rate correlates with the Reynolds number consistent with naive expectations, lending additional credibility to our interpretation of the exponential amplification of the magnetic field strength in terms of a turbulent dynamo. We note that it is not a priori clear if we should expect this result, because we are comparing many different haloes, on top of large uncertainties in the estimates of both the Reynolds number and the amplification rate. Moreover, the amplification rate even at our highest resolution simulations is much smaller than expected in reality, because the numerical resistivity in our simulations is much larger than realistic values for the physical resistivity \citep{Schober2013}.

Lastly we show $t_\mathrm{dynamo}$ in the bottom left panel of Figure~\ref{fig:bfld_amplification_overview}, as well as an estimate of the time when the dynamo actually started in the bottom right panel of Figure~\ref{fig:bfld_amplification_overview}. We obtain this estimate by extrapolating from $t_\mathrm{dynamo}$ backwards in time to $B(t)=B_0$ using the growth rate estimates shown in the top right panel of the same figure. The starting times for the dynamo amplification are of order $1\,\mathrm{Gyr}$ and do not change monotonically with resolution for individual haloes. We conclude that they are consistent with each other within the uncertainties of the estimate.

Therefore we argue that the onset of the turbulent dynamo is set by the evolution of the halo, i.e. its growth and early star formation history that generates outflows, rather than numerical limitations. This is not completely surprising because all our simulations reach Reynolds numbers $\mathrm{Re}\geq 100$, sufficient to support a turbulent dynamo according to idealised estimates \citep{Brandenburg2005,Schober2012} as well as practical experiences from numerical simulations of galaxies \citep{Martin-Alvarez2018}. In contrast, the amplification rate and therefore also the time when the magnetic field saturates, clearly depend on the numerical resolution of the simulations (see, e.g. Figure~\ref{fig:amplification_disc}).

\begin{figure}
    \centering
    \includegraphics[width=0.48\textwidth]{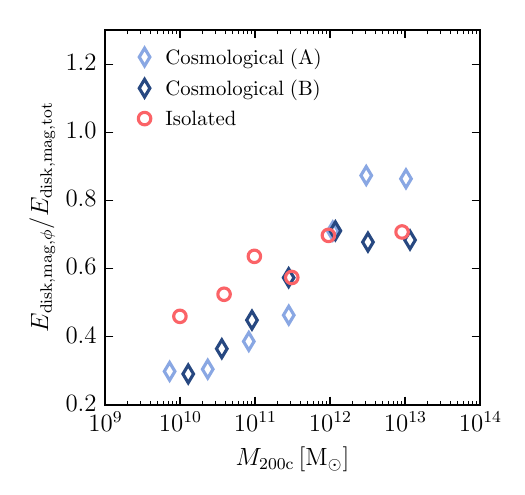}
    \includegraphics[width=0.48\textwidth]{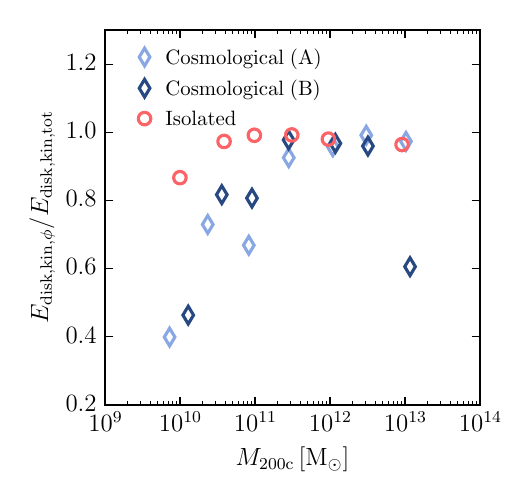}
    \caption{Fraction of magnetic energy (top panel) and kinetic energy (bottom panel) in the azimuthal component in comparison to the total magnitudes in the disc, respectively. The blue diamonds show all cosmological galaxies at $z=0$. Red circles give isolated galaxies by \citet{Jacob2018} at $3~\mathrm{Gyr}$. The magnetic field is dominated by its azimuthal component for galaxies in haloes with $\Mhalo\gtrsim 10^{11}~\mathrm{M_\odot}$. However, the azimuthal component of the magnetic field is always relatively less important than the azimuthal component of the velocity field.}
    \label{fig:bfld_structure}
\end{figure}

\section{Magnetic field structure at the present epoch}
\label{sec:structure}

We have, so far, focused on the strength of the magnetic field, which is a good measure of its dynamic impact. However, the structure of the magnetic field is equally important, in particular for transport processes along magnetic field lines, for example of cosmic rays \citep{Pakmor2016Diffusion,Thomas2021}, but also for various observational diagnostics of magnetic fields like Faraday rotation, or polarised radio emission \citep{Beck2015}.

As we demonstrated in Figure~\ref{fig:gas}, at $z=0$ our simulated galaxies have gas discs in the more massive haloes with $\Mhalo\geq 10^{11}\,\mathrm{M_\odot}$. We expect the rotation of the discs to order the magnetic field over time. In Figure~\ref{fig:bfld} we show the structure and strength of the magnetic fields in the gas discs at $z=0$. The galaxies in haloes with $\Mhalo\geq 10^{11}\,\mathrm{M_\odot}$ display large-scale, azimuthally ordered magnetic fields, whereas the two galaxies in less massive haloes do not, consistent with the absence of a dominant gas disc.

To get a better idea of the ordering process of the magnetic field we consider maps of the azimuthal component of the magnetic field of the cosmological galaxies at $z=0$ in the upper panels of Figure~\ref{fig:bphi}. Except for the galaxy in the $10^{12.5}\,\mathrm{M_\odot}$ halo, all cosmological galaxies have azimuthal magnetic fields with at least one field reversal. Although the detailed structure of the field reversals is highly stochastic, there is a general trend that galaxies in more massive haloes have a more ordered field with fewer reversals inside the disc, compared to galaxies in less massive haloes. Interestingly, the galaxies that feature a large scale ordered field with few reversals (in haloes $\Mhalo\geq 10^{11.5}\,\mathrm{M_\odot}$) are the same galaxies that reach equipartition between turbulent and magnetic energy (as shown in Figure~\ref{fig:energy_disc}).

For comparison we show the azimuthal component of the isolated galaxies at $3\,\mathrm{Gyr}$ in the lower panels of Figure~\ref{fig:bphi}. They exhibit very similar properties and trends. In particular the magnetic field in the disc is also more ordered with fewer reversals in more massive halos. The magnetic fields show clear signs of spiral structures. However, we postpone a detailed analysis of these spiral structures and correlations with other disc structures (such as magnetic and stellar spiral arms) and an in-depth study of the ordering process of the magnetic field in the discs to future work.

To quantify the importance of the azimuthal component of the magnetic field relative to its radial and vertical components at $z=0$ we look at its energy relative to the total magnetic energy in the volume of the disc in the upper panel of Figure~\ref{fig:bfld_structure}. For galaxies in haloes with $\Mhalo\geq 10^{12}\,\mathrm{M_\odot}$, the magnetic field in the disc is strongly dominated by the azimuthal component (i.e.~it contains at least $50\%$ of the magnetic energy). This is in contrast to the smallest galaxies in haloes of $\Mhalo\leq 10^{10.5}\,\mathrm{M_\odot}$, where the magnetic field has no preference for the azimuthal direction. This is again consistent with the absence of a rotationally supported gas disc at late times. Between these two extremes the importance of the azimuthal magnetic field steadily increases with increasing halo mass.

Figure~\ref{fig:bfld_structure} compares the relative importance of the azimuthal magnetic field in cosmological galaxies with isolated galaxies and highlights some interesting differences. Lower mass isolated galaxies in haloes of $\Mhalo\leq 10^{11}\,\mathrm{M_\odot}$ have a more important azimuthal magnetic field component in comparison to cosmological galaxies in haloes of the same mass. This is likely a direct consequence of rotation being more important for the isolated galaxies, as can be seen in the lower panel of Figure~\ref{fig:bfld_structure} which shows the fraction of kinetic energy in rotation over the total kinetic energy in the gas disc. Here we see quantitatively that in cosmological galaxies in haloes with $\Mhalo\leq 10^{11}\,\mathrm{M_\odot}$ rotation becomes less important for the dynamics of the gas disc, even though we already the galaxies to be the most discy of their halos mass range. In contrast, the isolated galaxies remain rotationally dominated even in the lowest mass haloes. This might motivate a smaller value for the initial spin parameter of these isolated haloes in order to make them more realistic in future simulations.

\section{Summary and outlook}
\label{sec:summary}

In this paper we have shown that in all our cosmological zoom simulations with sufficient numerical resolution a turbulent dynamo efficiently amplifies the magnetic field. It saturates already at high redshift and the saturation strength is converged in galaxies at all halo masses. The dynamo typically saturates when the magnetic energy density reaches $\gtrsim 10\%$ of the turbulent kinetic energy density. The time when saturation is reached depends strongly on numerical resolution. The dynamo saturates early enough only with sufficient resolution. Importantly, only then the galaxy has a physically meaningful magnetic field for most of its evolution.

The galaxies in the lowest mass haloes saturate below that value, possibly because of the shortcomings of our ISM subgrid model for the smallest galaxies. To better understand magnetic fields in the smallest dwarf galaxies and to quantify the dependence on the subgrid model we plan to compare our results to simulations of dwarf galaxies with a more detailed, explicit ISM model \citep[e.g. the Lyra model;][]{Gutcke2022} in the future.

The numerical resolution that is required to reach converged saturated magnetic fields depends strongly on halo mass. We need better than $10^3\,\mathrm{M_\odot}$ gas mass resolution for the smallest $10^{10}\,\mathrm{M_\odot}$ halo, a gas resolution of $5\times 10^{4}\,\mathrm{M_\odot}$ or better for haloes with $10^{11}\,\mathrm{M_\odot}\leq \Mhalo\leq 10^{11.5}\,\mathrm{M_\odot}$, a gas resolution of $4\times 10^{5}\,\mathrm{M_\odot}$ or better for Milky Way-like galaxies with a halo mass of $\Mhalo\sim 10^{12}\,\mathrm{M_\odot}$, and gas resolution of $3\times 10^{6}\,\mathrm{M_\odot}$ for more massive haloes with $\Mhalo\geq 10^{12.5}\,\mathrm{M_\odot}$. This very roughly translates to a minimum of $10^6$ gas cells (of equal mass) in the whole halo. 

Note that these numbers certainly depend on the numerical scheme as well as on the feedback model used. They are therefore only directly applicable to the Auriga model, but should still provide a good guideline for the IllustrisTNG model as well due to its close similarity.

More fundamentally though, our understanding of MHD turbulence is still incomplete and MHD turbulence is currently an area of active research. Analytical studies and numerical simulations indicate that the saturation level of the magnetic field, i.e.~the ratio of magnetic to kinetic energy at saturation, depends significantly on the Prandtl number, i.e.~the ratio between the magnetic and the kinetic Reynolds numbers \citep{Federrath2014,Schober2015,Kriel2023}. For Prandtl numbers significantly larger than unity, as expected for the volume filling warm phase of the interstellar medium \citep{Rincon2019}, the ratio between magnetic and kinetic energy can even reach unity. This regime, however, is currently still impossible to reach in galaxy simulations because of a lack of numerical resolution. Moreover, in addition to numerical resolution, different models for the interstellar medium and feedback processes \citep{Springel2003, Grand2016, Rieder2017, Su2018, Hopkins2019, Martin-Alvarez2022} as well as the numerical scheme \citep{Teyssier2006, Pakmor2013, Mocz2016, Hopkins2019} are important ingredients that could influence the evolution of magnetic fields in simulations of galaxies.

We demonstrated that for haloes with masses $\Mhalo\geq 10^{11}\,\mathrm{M_\odot}$ the circumgalactic medium is also fully magnetised at $z=0$ at a level of $\sim 10\%$ of equipartition between magnetic and kinetic energy. The magnetisation of the halo is started by magnetised outflows from the galaxy, similar to previous results for Milky Way-like galaxies \citep{Pakmor2020}. In lower mass haloes, the CGM is essentially not magnetised as galactic outflows do not penetrate far enough into the halo in the Auriga model, and their CGM is not turbulent enough to support an in-situ dynamo.

Our cosmological galaxies in haloes with $\Mhalo\geq 10^{12}\,\mathrm{M_\odot}$ are dominated by azimuthal magnetic fields. This dominance weakens with decreasing halo mass and is absent for the smallest dwarf galaxies in haloes of mass $\Mhalo\leq 10^{10.5}\,\mathrm{M_\odot}$.

We compare our galaxies to simulations of isolated galaxies by \citet{Jacob2018}. These isolated galaxies from the collapsing halo setup are a surprisingly good representation of low-redshift cosmological galaxies in haloes of $\Mhalo\geq 10^{11.5}\,\mathrm{M_\odot}$, despite lacking most of their accretion history and the gradual buildup of stellar mass. Lower mass isolated galaxies have weaker magnetic fields than cosmological galaxies in haloes of the same mass, because they are significantly less turbulent. Ordered rotation is also more important in low mass isolated galaxies compared to our cosmological simulations.

In the future, we plan to examine in more detail the magnetic fields in the CGM, their dynamical effects, and how those change with halo mass. Having shown convergence for the magnetic fields, we will simulate the galaxies studied in this paper augmented with cosmic rays \citep{Buck2019} to investigate their effect on the gas flows in and around galaxies in a cosmological environment for a wide range of halo masses. Reliable magnetic fields will also allow us to include other physical transport processes that depend on the structure of the magnetic field, such as anisotropic thermal conduction \citep{Kannan2016}.

Magnetic fields are finally becoming a more commonly included component in cosmological galaxy simulations \citep[see, e.g.][]{Pakmor2014,Pakmor2017,Hopkins2019,Martin-Alvarez2022,Liu2022}. They already allow us to routinely generate various synthetic observables connected to magnetic fields and compare them to observations \citep{Pakmor2018, Ponnada2022, Jung2023, Reissl2023, Martin-Alvarez2023, Ponnada2023}. However, we are still only scratching the surface of many questions and it needs to be checked whether our results hold in other simulation codes and with different galaxy formation models.

Isolated galaxy simulations, which allow for much higher resolution, and therefore an easier inclusion of more physical processes, and also provide a more controlled way to understand specific questions and setups \citep[see, e.g.][]{Jacob2018,Bieri2022,Pfrommer2022,Thomas2023}, will remain an important approach to improve our understanding of how galaxies evolve. Such simulations can likely be improved and made more realistic in the future by using cosmological simulations to motivate the choice of initial (and boundary) conditions. At the same time, results from simulations of high resolution isolated galaxies can lead to improvements of subgrid models in cosmological simulations. We conclude that combining high resolution cosmological zoom simulations with even higher resolution simulations of isolated galaxies holds a lot of promise to understand magnetic fields and other physical processes connected to and depending on them.

\section*{Acknowledgements}

We thank the referee for their constructive and helpful comments that improved the paper. RB acknowledges support from the exoclimes simulation platform (Projekte MINT) Nr. 200020\_192022. FvdV is supported by a Royal Society University Research Fellowship (URF\textbackslash R1\textbackslash 191703). CP acknowledges support by the European Research Council under ERC-AdG grant PICOGAL-101019746. AF is supported by a UKRI Future Leaders Fellowship (grant no. MR/T042362/1). 

The authors gratefully acknowledge the Gauss Centre for Supercomputing e.V. (https://www.gauss-centre.eu) for funding this project (with project code pn68ju) by providing computing time on the GCS Supercomputer SuperMUC-NG at Leibniz Supercomputing Centre (https://www.lrz.de). This work used the DiRAC@Durham facility (under project code dp221) managed by the Institute for Computational Cosmology on behalf of the STFC DiRAC HPC Facility (www.dirac.ac.uk). The equipment was funded by BEIS capital funding via STFC capital grants ST/K00042X/1, ST/P002293/1, ST/R002371/1 and ST/S002502/1, Durham University and STFC operations grant ST/R000832/1. DiRAC is part of the National e-Infrastructure.

\section*{Data Availability}
The data underlying this article will be shared on reasonable request to the corresponding author.



\bibliographystyle{mnras}





\bsp	
\label{lastpage}
\end{document}